%% file: main.tex
\documentclass[sigconf, nonacm]{acmart}
\input{header}

\newcommand\vldbdoi{XX.XX/XXX.XX}
\newcommand\vldbpages{XXX-XXX}
\newcommand\vldbvolume{14}
\newcommand\vldbissue{1}
\newcommand\vldbyear{2020}
\newcommand\vldbauthors{\authors}
\newcommand\vldbtitle{\shorttitle} 
\newcommand\vldbavailabilityurl{URL_TO_YOUR_ARTIFACTS}
\newcommand\vldbpagestyle{plain} 

\begin{document}
\title{\sys: A Practical Incentive-Aware Federated Learning Framework}
%
\author{Peishen Yan}
\authornote{Equal contribution.}
\affiliation{%
  \institution{Shanghai Jiao Tong University}
}
\email{peishenyan@sjtu.edu.cn}

\author{Shuang Liang}
\authornotemark[1]
\affiliation{%
  \institution{Shanghai Jiao Tong University}
}
\email{liangshuangde@sjtu.edu.cn}

\author{Yang Hua}
\affiliation{%
  \institution{Queen's University Belfast}
}
\email{Y.Hua@qub.ac.uk}

\author{Linshan Jiang}
\affiliation{%
  \institution{National University of Singapore}
}
\email{linshan@nus.edu.sg}

\author{Kuai Yu}
\affiliation{%
  \institution{Columbia University}
}
\email{ky2589@columbia.edu}

\author{Yulin Sun}
\affiliation{%
  \institution{Shanghai Jiao Tong University}
}
\email{rain-forest@sjtu.edu.cn}

\author{Yaozhi Zhang}
\affiliation{%
  \institution{Shanghai Jiao Tong University}
}
\email{zhangyaozhi@sjtu.edu.cn}

\author{Tao Song}
\affiliation{%
  \institution{Shanghai Jiao Tong University}
}
\email{songt333@sjtu.edu.cn}

\author{Ningxin Hu}
\affiliation{%
  \institution{Intel Corporation}
}
\email{ningxin.hu@intel.com}

\author{Xinran Liang}
\affiliation{%
  \institution{United Imaging Intelligence Co., Ltd}
}
\email{xinran.liang@uii-ai.com}

\author{Bingsheng He}
\affiliation{%
  \institution{National University of Singapore}
}
\email{dcsheb@nus.edu.sg}

\author{Haibing Guan}
\authornote{Corresponding author.}
\affiliation{%
  \institution{Shanghai Jiao Tong University}
}
\email{hbguan@sjtu.edu.cn}
\renewcommand{\shortauthors}{Yan \etal}
%
\input{sec/0_abstract}
\maketitle

\pagestyle{\vldbpagestyle}
\begingroup\small\noindent\raggedright\textbf{PVLDB Reference Format:}\\
\vldbauthors. \vldbtitle. PVLDB, \vldbvolume(\vldbissue): \vldbpages, \vldbyear.\\
\href{https://doi.org/\vldbdoi}{doi:\vldbdoi}
\endgroup
\begingroup
\renewcommand\thefootnote{}\footnote{\noindent
This work is licensed under the Creative Commons BY-NC-ND 4.0 International License. Visit \url{https://creativecommons.org/licenses/by-nc-nd/4.0/} to view a copy of this license. For any use beyond those covered by this license, obtain permission by emailing \href{mailto:info@vldb.org}{info@vldb.org}. Copyright is held by the owner/author(s). Publication rights licensed to the VLDB Endowment. \\
\raggedright Proceedings of the VLDB Endowment, Vol. \vldbvolume, No. \vldbissue\ %
ISSN 2150-8097. \\
\href{https://doi.org/\vldbdoi}{doi:\vldbdoi} \\
}\addtocounter{footnote}{-1}\endgroup

\ifdefempty{\vldbavailabilityurl}{}{
\vspace{.3cm}
\begingroup\small\noindent\raggedright\textbf{PVLDB Artifact Availability:}\\
The source code, data, and/or other artifacts have been made available at \url{https://github.com/wizicer/web3fl}.
\endgroup
}

\input{sec/1_intro}
\input{sec/2_related}
\input{sec/3_design}
\input{sec/4_impl}
\input{sec/5_exp}
\input{sec/6_conclusion}

\clearpage

\bibliographystyle{ACM-Reference-Format}
\bibliography{main}

\input{sec/appendix}
\end{document}

%% file: header.tex
\usepackage{textcomp}
\usepackage{color}
\usepackage{xcolor}
\usepackage[ruled,linesnumbered]{algorithm2e}
\usepackage{multirow}
\usepackage{makecell}
\usepackage{booktabs}
\usepackage{outlines}
\usepackage{bbold}
\usepackage{subfigure}
\usepackage{enumitem}
\usepackage{soul}
\usepackage{color}
\usepackage{comment}
\usepackage{diagbox}
\usepackage{pifont}
\usepackage{textcomp}
\usepackage{lipsum}
\usepackage{longtable}

\definecolor{ForestGreen}{RGB}{34,139,34}

\definecolor{darkgreen}{rgb}{0.0, 0.5, 0.0}

\newcommand{\eg}{{\it e.g.}}
\newcommand{\etal}{\textit{et al}.~}

\newcommand{\ie}{{\it i.e.}}

\newcommand{\sys}{\textsc{FWeb3}\xspace}

%% file: sec/0_abstract.tex
\begin{abstract}
Federated learning (FL) enables collaborative model training over distributed private data. However, sustaining open participation requires incentive mechanisms that compensate contributors for their resources and risks. 
Enabled by Web3 primitives, especially blockchains, recent FL proposals incorporate incentive mechanisms for open participation, yet most focus primarily on algorithmic design and overlook system-level challenges, including coordination efficiency, secure handling of model updates, and practical usability.
We present \sys, a practical Web3-enabled FL framework for incentive-aware training in open environments. \sys adopts a modular architecture that separates FL functions from Web3 support services, decoupling the off-chain training and data plane from on-chain settlement while preserving verifiable incentive execution. The framework supports pluggable aggregation and contribution evaluation methods and provides a browser-native DApp interface to lower the participation barrier.
We evaluate \sys in real-world settings and show that it supports end-to-end incentive-aware FL with transaction and data-transfer overheads of only 21.3\% and 3.4\% in WAN; \sys also deploys from zero configuration in under 3 minutes and enables user onboarding in under 1 minute.
\end{abstract}

%% file: sec/1_intro.tex
\section{Introduction}\label{sec:intro}
Modern machine learning applications increasingly rely on large volumes of high-quality data, posing growing challenges for data management. While public datasets have driven early progress, many real-world applications in healthcare and finance depend primarily on private, user-owned data that cannot be centrally collected or freely shared~\cite{meurisch2021data,oldenhof2023industry,cao2022ai}.

When valuable data is fragmented across many owners and cannot be centralized, collaborative learning requires a mechanism to organize distributed private data into a shared computation workflow. From a data management standpoint, federated learning (FL) is not merely a learning algorithm but a primitive for organizing distributed private data into a recurring computation-and-aggregation process~\cite{kairouz2021advances}. In principle, this could enable an open data ecosystem where data owners contribute locally and continuously improve shared models, provided that open participation is sustainable.

The main obstacle to sustaining this vision is incentives: in open environments, participants bear non-trivial resource costs and potential privacy or compliance exposure~\cite{zhan2021survey,zhu2019deep}, yet lack a principled and enforceable compensation mechanism.
As a result, incentive-free FL largely remains limited to small, pre-arranged collaborations. Although many recent systems add incentives, they often treat them as an algorithmic afterthought, leaving system-level challenges underexplored~\cite{yu2020fairness,weng2019deepchain}.
Consequently, existing deployments either require permissioned participation, incur substantial coordination overhead, or impose heavyweight client-side environments, limiting their applicability in real-world data ecosystems.
Overall, supporting incentives in FL is a system design problem: it requires rethinking settlement, dataflow, security, and usability under open participation. We next summarize four system-level requirements that an incentive-compatible FL system must satisfy in such environments.

\noindent\textbf{R1. Transparent and Permissionless Incentive Settlement.} Introducing incentive mechanisms turns FL into an economic system that requires transparent and auditable settlement across geo-distributed participants. In open environments, permissionless settlement with verifiable reward execution accounting better supports broad participation, allowing data owners to engage without centralized authorization. However, conventional cross-organizational payment infrastructures rely on centralized intermediaries and pre-established trust, and may involve delayed settlement due to reconciliation and compliance procedures, making them ill-suited for open and scalable incentive-compatible FL. Consequently, practical incentive-compatible FL calls for a settlement layer that supports open participation with instant, verifiable, and programmable reward execution. \emph{Web3 primitives, such as blockchains and smart contracts, provide a practical foundation for such settlement in open environments}~\cite{nguyen2021federated}.

\noindent\textbf{R2. Systematic Efficiency under Web3 Constraints.} Integrating blockchain-based settlement and coordination into FL directly alters the dataflow of model updates and control messages. Naively replacing components in the traditional FL pipeline with Web3 primitives, or excessively interacting with the blockchain along the training path, introduces substantial coordination and communication overhead, which can significantly slow down training rounds. Therefore, practical incentive-aware FL should better co-design its dataflow to bound the overhead introduced by blockchain interaction and preserve the efficiency of standard FL workflows.

\noindent\textbf{R3. Security of Model Updates for Incentive Integrity.} Permissionless economic participation fundamentally changes the threat model of FL. In open environments, behaviors such as free-riding, unauthorized reuse of model updates, or model theft can directly distort contribution evaluation and reward allocation. As incentives are tied to model updates, protecting the confidentiality and integrity of local models and intermediate updates becomes essential for maintaining correct incentive execution.

\noindent\textbf{R4. Usability and Extensibility for Real Adoption.} To enable real-world adoption, incentive-compatible FL systems must remain accessible to non-expert users and extensible for developers. Data owners should be able to join and leave dynamically without specialized machine learning or blockchain expertise, while developers should be able to integrate new aggregation or contribution-evaluation algorithms without re-engineering the underlying Web3 communication and security infrastructure.

These requirements abstract incentive-compatible FL into four system-level dimensions, each of which addresses a failure mode observed in existing systems. Omitting any one of them often leads to impractical deployments, while additional considerations typically reduce to refinements within these dimensions. Together, they delineate the practical design space for incentive-compatible FL under open participation

In this paper, we present \sys, a practical Web3-enabled FL framework designed for open, incentive-aware environments. At its core, \sys leverages blockchain as a transparent settlement layer to record contribution outcomes and execute reward distribution through smart contracts, eliminating reliance on trusted intermediaries and enabling permissionless participation (\textbf{R1}).

To preserve training-path efficiency, \sys decouples FL logic from blockchain execution. It adopts a modular architecture that separates FL functionality from Web3 supporting layer, decomposing the workflow into four functional modules (\textit{configuration}, \textit{training}, \textit{aggregation}, and \textit{contribution evaluation}) organized over three supporting components (\textit{trusted transaction}, \textit{information communication}, and \textit{information protection}).
Through a hybrid data plane and minimal on-chain interaction, models and messages are routed off-chain, while the blockchain is used only for essential commitments and metadata, thereby bounding coordination overhead and preserving the efficiency of standard FL workflows (\textbf{R2}).

To safeguard model updates under open participation, \sys co-designs security with the model-exchange dataflow.
It enforces secure access control and combines lightweight cryptographic mechanisms to ensure confidentiality and integrity of exchanged updates, preventing unauthorized reuse or model theft that could compromise contribution evaluation and reward allocation (\textbf{R3}).

Finally, \sys provides a browser-native decentralized application (DApp) interface to lower the barrier to participation, while its modular architecture enables developers to integrate new algorithms without re-engineering the underlying Web3 communication and security components (\textbf{R4}).

In summary, our key contributions are as follows:

\begin{itemize}[leftmargin=1em]
\item We abstract incentive-compatible federated learning in open environments into four system-level requirements---\emph{transparent and permissionless incentive settlement}, \emph{systematic efficiency under decentralization constraints}, \emph{model-update security for incentive integrity}, and \emph{usability and extensibility for real adoption}. These requirements together delineate the practical design space of incentive-compatible FL under open participation.

\item Guided by the above requirements, we design and implement \sys, a practical Web3-enabled federated learning framework with a modular architecture that separates federated learning execution from Web3 support services. \sys provides a web-native DApp interface and stable extension points for aggregation and contribution evaluation, enabling low-friction participation and lightweight algorithm integration.

\item Simulations and real-world experiments in LAN and WAN settings show that \sys enables end-to-end incentive-aware federated learning with acceptable non-training overhead, while supporting practical deployment and user onboarding in open-participation settings.
\end{itemize}

%% file: sec/2_related.tex
\section{Background}\label{sec:background}

\subsection{Federated Learning}

A typical FL system configuration comprises a server and a network of $n$ clients. Each client, indexed as $i$, possesses its localized dataset denoted as $D_i$, where $i$ ranges from 1 to $n$. Within each communication round $t$, the server strategically selects a subset of clients, represented as $\mathcal{N}_t$, and orchestrates a sequence of actions: 1) \emph{Model distribution:} Initially, the server distributes the global model, referred to as $W_t$, to the chosen set of clients. 2) \emph{Local training:} Clients, upon receipt of the global model $W_t$, initiate local training by considering $W_t$ as their initial model $W^{(i)}_{t+1}$. They iteratively perform updates using the expression $W^{(i)}_{t+1}:=W^{(i)}_{t+1}-\beta \nabla f(W^{(i)}_{t+1};D_i)$ for a predefined number of local iterations $l$. In this equation, $f(\cdot;\cdot)$ represents the empirical loss function, and $\beta$ signifies the local learning rate. The corresponding local model update is captured as $\Delta W^{(i)}_{t+1}=W_t-W^{(i)}_{t+1}$, effectively equating to the local model $W^{(i)}_{t+1}$ from the server's perspective. 3) \emph{Model aggregation:} Subsequent to local training, each client $i$ uploads its model weights, denoted as $W^{(i)}_{t+1}$, to the server for aggregation. The well-known FedAvg algorithm~\cite{mcmahan2017communication} employs weighted model averaging to update the global model: $W_{t+1}=\sum_{i\in \mathcal{N}_t} \frac{|D_i|}{\sum_{i\in \mathcal{N}_t}|D_i|}W^{(i)}_{t+1}$. 

To promote sustainability in FL, modern FL systems often include a client \emph{contribution evaluation} stage~\cite{chen2024contribution}. After aggregation, the server assesses each client's update $\Delta W^{(i)}_{t+1}$ based on its contribution on the global model. Using an evaluation metric $\mathcal{M}(\cdot)$ (\eg, accuracy gain or loss reduction), the contribution of client $i$ in round $t$ is defined as $c^{(i)}_t = \mathcal{M}(W_{t+1}, W_{t}, \Delta W^{(i)}_{t+1})$. The resulting scores guide incentive allocation, encouraging participation and fostering a sustainable FL ecosystem.

The above contribution evaluation stage naturally turns FL from a purely collaborative training protocol into an \emph{economic} process: participants contribute data and computation, and the system must translate their measured utility into rewards.
This setting motivates \emph{incentive-aware FL}, which studies how to design reward and participation rules so that providing high-quality updates is a rational strategy for self-interested participants~\cite{sun2024hifi,zhang2021incentive,wang2025dealing}.
Existing efforts typically follow two complementary directions.
One direction focuses on \emph{mechanism design} for participation and payment, drawing on auction-based selection, contract-theoretic modeling under information asymmetry, and game-theoretic formulations (\eg, Stackelberg leader-follower) to regulate client effort and reward allocation~\cite{gao2021fifl,wang2023incentive,wang2024fast,tang2025game}.
Another direction focuses on \emph{contribution accounting}, estimating each client's marginal utility (\eg, via Shapley-style or utility-based approximations) to support fair reward allocation~\cite{zheng2023secure,wei2025efficient}.
However, most prior studies primarily operate at the algorithmic level and leave system concerns, including auditable settlement, low-overhead execution along the training path, and security under open participation, largely unexplored. This gap motivates the need for a practical incentive-aware FL system framework.

\subsection{Incentive-aware FL Design Considerations}

We analyze the system-level challenges of incentive-aware FL through reported statistics, analysis, and pilot deployments. We focus on four key dimensions.

\subsubsection{Transparent and Permissionless Incentive Settlement}
A practical incentive mechanism requires not only how rewards are computed, but also how they are \emph{settled} in a verifiable and low-friction manner under open participation.
Existing FL platforms highlight the limitations of current settlement models.
Existing FL software stacks primarily focus on orchestrating distributed training, and typically assume a pre-arranged participant set governed by organizational agreements, identity provisioning, and security policies, rather than providing built-in, verifiable incentive settlement.
As representative examples, FedML and Flower~\cite{he2020fedml,beutel2020flower} facilitate experimentation and deployment but do not include native reward settlement, while OpenFL~\cite{reina2021openfl} targets consortium-style collaboration and relies on pre-established coordination and security configurations among participating organizations, which limits openness and frictionless cross-organizational participation.
These observations suggest that sustaining open FL requires a settlement mechanism that is both \emph{transparent} and \emph{permissionless}.

In practice, incentive settlement in FL often corresponds to real economic transactions across organizations or jurisdictions.
Industry reports~\cite{reuter2025payment,bindseil2022towards} indicate that traditional cross-border payments typically involve multiple intermediaries and compliance procedures, leading to settlement delays of one to several business days and non-trivial transaction fees.
For example, traditional wire transfers can take 1–5 business days to settle due to routing through multiple banks and compliance checks~\cite{netsuite_crossborder_2025,worldbank_remittance_2024}. Moreover, globally, sending remittances costs on average about 6.49\% of the amount transferred, reflecting the combined effect of transfer fees, intermediary charges, and exchange spreads~\cite{reutersG202025}.
Participation further requires formal account registration and contractual arrangements, which impose significant entry barriers for small or transient contributors.
These properties make fine-grained and frequent reward distribution difficult to execute transparently and efficiently in open settings.
However, Web3 primitives, such as blockchains and smart contracts, provide a natural foundation for programmable and auditable settlement, enabling participants to join and receive rewards without relying on centralized authorization or trusted intermediaries.
On public blockchain networks such as Ethereum, typical on-chain transactions can be confirmed in under a minute~\cite{ethereum_stats_2025}, and average transaction fees remain relatively low (\eg, around \$0.20–\$0.33 per transaction in late 2025)~\cite{ethereum_gas_fees_2025}, illustrating how decentralized settlement habitats can support frequent, verifiable execution at scale.

\subsubsection{Systematic Efficiency under Web3 Constraints}

A common misconception is that introducing blockchain for incentive settlement implies moving the entire FL workflow on-chain. In practice, blockchains are primarily used for accountability and reward execution, while learning itself must remain off-chain due to fundamental system constraints.

Prior blockchain-based FL proposals often highlight immutability and auditability as key advantages~\cite{ying2024bitfl,fan2020hybridbfl}, but overlook the efficiency limits of public chains. Public blockchains impose hard bounds on the frequency and complexity of on-chain execution, due to block generation rate, block capacity (\eg, gas limits), and limited throughput~\cite{wood2014ethereum}. We summarize representative public-chain execution constraints in Table~\ref{tab:blockchain_performance}.

\begin{table}[t]
\centering
\small
\caption{Representative performance constraints of major public blockchains. TPS denotes transactions per second.}
\label{tab:blockchain_performance}
\setlength{\tabcolsep}{1mm}
\begin{tabular}{lcccc}
\toprule
\textbf{Blockchain} & \textbf{Block Time} & \textbf{TPS} & \textbf{Hardware Barrier} & \textbf{Stability} \\
\midrule
Bitcoin  & $\sim$10 min  & 3--7   & Low  & High \\
Ethereum (L1) & $\sim$12 s & 25--30 & Low--Medium & High \\
Solana   & $\sim$0.4 s  & $>$1,400 & High & Medium \\
\bottomrule
\end{tabular}
\end{table}

Meanwhile, modern neural network training and aggregation are dominated by compute- and memory-intensive primitives (\eg, GEMM, convolution, backpropagation, and large-scale reduction), whose operation counts and memory footprints scale with model size.
Recent studies on on-chain machine learning show that even modest neural inference or training workloads translate into millions of gas-consuming operations, far exceeding practical block execution budgets~\cite{alhaidari2025chain,wood2014ethereum}.
As a result, even a single forward or backward pass involves orders of magnitude more computation than what can be feasibly executed on-chain, rendering direct on-chain learning or aggregation impractical.

Motivated by these limits, recent systems decouple learning from blockchain consensus: training and aggregation are performed off-chain, while the chain is used for state commitment and settlement~\cite{lin2024refiner}. However, decoupling computation alone is insufficient. A practical Web3-enabled FL system must handle the dataflow of model updates and control messages across Web3 primitives (smart contracts and decentralized storage). Naively uploading model updates to distributed storage (\eg, IPFS) and recording only the CID on-chain does not address transmission latency, availability variance, or bandwidth contention under realistic network settings.

\begin{figure}
    \centering
    \includegraphics[width=0.85\linewidth]{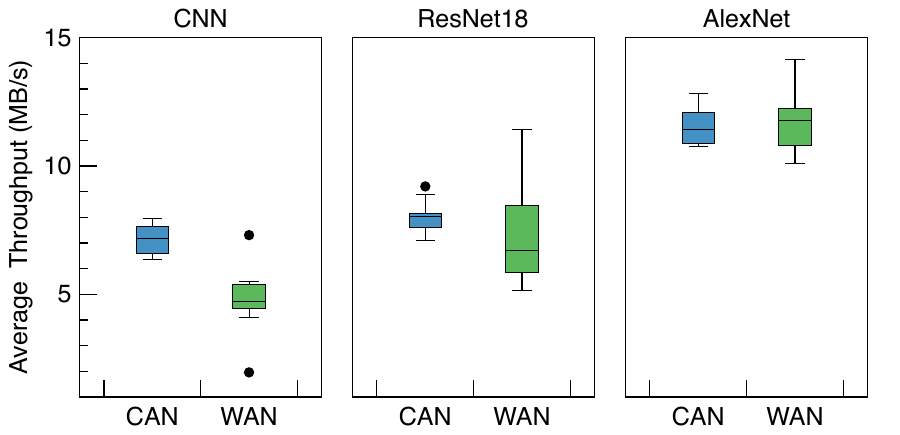}
    \caption{End-to-end IPFS transfer throughput for exchanging model updates under CAN and WAN settings.}
    \label{fig:ipfs-profiling}
\end{figure}

\noindent\textbf{Pilot Observation.} To characterize the communication bottleneck of IPFS-based model exchange, we conduct a two-device microbenchmark that measures the transfer throughput of model updates via IPFS. We evaluate three representative model sizes, including a 4-layer CNN (5.6\,MB), ResNet18 (44.7\,MB), and AlexNet (233\,MB). We run the benchmark in a campus-area network (CAN), where the two devices are located in the same region, and a wide-area network (WAN), where the devices are placed in different regions within the same continent. Figure~\ref{fig:ipfs-profiling} shows that IPFS throughput is sensitive to network conditions, and WAN generally reduces throughput and increases variance compared to CAN.
This empirical bottleneck motivates a hybrid update dissemination design (\S\ref{sec:datatrans}), where direct peer-to-peer delivery is preferred and IPFS is used as a best-effort fallback rather than the primary data plane.

These observations also suggest that systematic efficiency in Web3-enabled FL requires more than moving computation off-chain. Practical frameworks should treat public blockchains as a lightweight settlement and audit layer rather than an execution substrate, and co-design the control-plane interactions with an efficient off-chain data plane for model updates. This co-design is essential to bound coordination and transmission overhead while preserving security and correctness in open, Web3-enabled FL systems.

\subsubsection{Security of Model Updates for Incentive Integrity}

The introduction of economic incentives fundamentally alters the threat model of federated learning. Beyond classical concerns such as privacy leakage~\cite{melis2019exploiting} and Byzantine robustness~\cite{blanchard2017machine}, incentive-driven FL must account for \emph{strategic} behaviors that directly target reward allocation.
Unlike classical adversaries that aim to degrade model utility~\cite{bagdasaryan2020backdoor}, self-interested participants may seek to maximize rewards while minimizing their own computation or data contribution~\cite{kang2019incentive,gao2021fifl}.
This rational incentive structure naturally induces behaviors such as free-riding, update replay, and contribution imitation~\cite{yu2023ironforge}, which distort contribution evaluation and lead to unfair incentive distribution.

In permissionless environments, model updates become economically valuable artifacts. Without adequate protection, malicious participants may reuse or copy others' updates, submit fabricated gradients, or exploit weaknesses in evaluation protocols to obtain rewards without performing meaningful computation~\cite{fraboni2021free}.
Notably, such attacks can succeed even when the incentive settlement logic itself is correctly implemented and transparently executed, as they manipulate the inputs to reward allocation rather than the allocation rules.

As a result, incentive integrity depends not only on correct reward execution, but also on the authenticity, integrity, and provenance of the model updates on which rewards are based~\cite{xu2019verifynet}.
Ensuring that each rewarded update corresponds to genuine local computation requires system-level mechanisms that bind model updates to their legitimate contributors and prevent unauthorized reuse or forgery~\cite{bonawitz2017practical}.
This necessitates the joint use of cryptographic authentication, secure communication, and controlled access to intermediate models~\cite{bell2020secure}, so that economic incentives remain aligned with real learning contributions under open participation.

\subsubsection{Usability and Extensibility for Real-world Adoption.}

Despite increasing research interest in blockchain-based incentive-aware FL, real-world adoption remains limited. Many existing systems require participants to manage cryptographic identities, configure wallets, interact directly with smart contracts, or deploy specialized runtime environments~\cite{ying2024bitfl,li2025veryfl}. These requirements impose significant technical and cognitive burdens on users, particularly non-expert data owners, and create high barriers to entry that reduce participation diversity.

From a system development perspective, tightly coupled designs further hinder extensibility. Integrating new aggregation algorithms, contribution evaluation methods, or incentive schemes often requires substantial re-engineering of the underlying blockchain and communication components. This lack of modularity slows innovation and limits the practical applicability of such frameworks across diverse use cases~\cite{kairouz2021advances}.

To support real-world deployment, incentive-aware FL systems must therefore abstract away Web3 complexity from end users while remaining modular and extensible for developers. Achieving this balance between accessibility and flexibility constitutes a key system-level challenge that is not captured by existing studies.

\input{tabs/new-taxonomy}
\subsection{Taxonomy of Open-Source FL Platforms and Incentive-Aware FL Prototypes}
\label{sec:taxonomy}

To contextualize incentive-aware FL from a system perspective, we organize representative implementations along two recurring archetypes:
(i) mature open-source FL platforms that emphasize training and deployment engineering, and
(ii) incentive-aware blockchain-based prototypes that incorporate decentralized reward execution.
Table~\ref{tab:taxonomy} summarizes how these systems cover the four system dimensions identified in previous section.

Overall, open-source FL platforms provide strong engineering support for orchestrating training and deployment, but typically externalize economic incentives and reward settlement to organizational agreements or off-system processes, rather than offering verifiable, programmable settlement as a first-class component.
In contrast, incentive-aware prototypes demonstrate the feasibility of using Web3 primitives for auditable settlement and accountability, but many remain proof-of-concept systems that do not provide a reusable, end-to-end abstraction across the settlement layer, the off-chain data plane, and security mechanisms.
As a result, they often incur non-trivial coordination overhead along the training path or impose significant deployment and integration effort when adapting to different deployment environments, which limits reuse and practical adoption at scale.

Taken together, the ecosystem lacks a reusable framework that simultaneously achieves transparent settlement, efficient training, secure model-update handling, and practical usability under open participation.
This gap motivates the design of \sys, which aims to unify these requirements within a modular Web3-enabled architecture.

\begin{figure*}[t]
    \centering
    \includegraphics[width=0.85\linewidth]{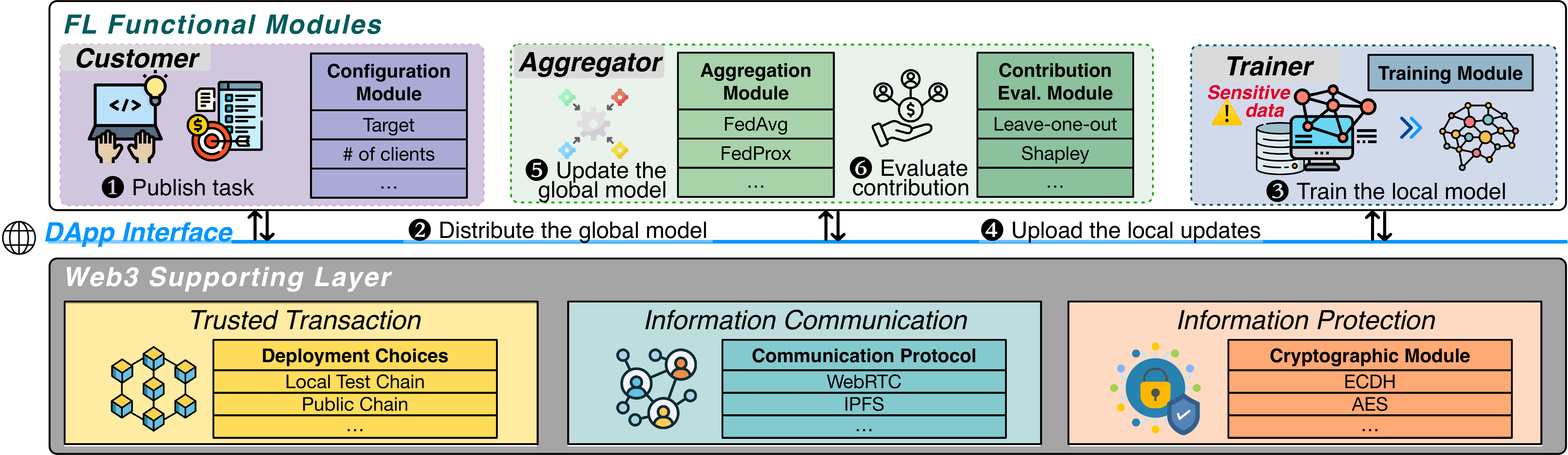}
    \caption{Overview of \sys. The framework involves three roles (customer, trainer, and aggregator) and four FL functional modules (configuration, training, aggregation, and contribution evaluation), organized over a Web3 supporting layer with three components (trusted transaction, information communication, and information protection).}
    \label{fig:overview}
\end{figure*}

%% file: tabs/new-taxonomy.tex
\begin{table*}[t]
\centering
\small
\caption{Taxonomy of representative federated learning systems. 
``(Un)supported'' indicates native (un)support within the framework, 
while ``Limited'' indicates partial or non-systematic support.}
\label{tab:taxonomy}

\begin{tabular}{lcccc}
\toprule
\textbf{System} 
& \makecell{\textbf{Transparent} \textbf{Settlement}} 
& \makecell{\textbf{Training-path} \textbf{Efficiency}} 
& \makecell{\textbf{Update} \textbf{Security}} 
& \makecell{\textbf{Usability \&} \textbf{Extensibility}} \\
\midrule
\textit{Open-Source FL Platforms}\\
\midrule
FedML~\cite{he2020fedml}      & Unsupported & High & Medium & High \\
Flower~\cite{beutel2020flower} & Unsupported & High & Medium & High \\
OpenFL~\cite{reina2021openfl}  & Unsupported & High & High & Medium \\
NVFlare~\cite{roth2022nvidia}   & Unsupported & High & High & High \\
FATE~\cite{liu2021fate}        & Unsupported & Medium & High & Medium \\
Substra~\cite{galtier2019substra}& Unsupported & Medium & High & Medium \\

\midrule
\textit{Incentive-Aware FL Prototypes}\\
\midrule
Refiner~\cite{lin2024refiner} & Limited & Low & Medium & Low \\
BlockDFL~\cite{fan2020hybridbfl} & Limited & Low & Medium & Low \\
BIT-FL~\cite{ying2024bitfl}      & Supported & Low & High & Low \\
BlockFLow~\cite{mugunthan2021blockflow} & Limited & Low & Low & Low \\
VeryFL~\cite{li2025veryfl} & Limited & Medium & Medium & Low \\
OpenFL~\cite{wahrstatter2024openfl} & Limited & Medium & Medium & Low \\

\midrule

\textbf{Ours} & Supported & High & High & High \\
\bottomrule
\end{tabular}
\end{table*}

%% file: sec/3_design.tex
\section{FWeb3 Design}

\subsection{Overview}

Figure~\ref{fig:overview} presents the overall architecture and workflow of \sys, which operationalizes federated learning as an incentive-aware and economically accountable collaborative process.
%
The process starts when a \textit{customer} issues an FL task on-chain, specifying training configurations such as the required number of clients, model type, and data format. \textit{Trainers} then register for the task, download the initialized global model, conduct local training on their private data, and submit model updates. One or more \textit{aggregators} collect these updates, refine the global model, and evaluate each contribution. 
%
The resulting contribution scores and settlement records are committed to the blockchain, where smart contracts automatically execute reward distribution according to predefined incentive rules.
Through this process, learning progress and economic settlement are tightly coupled yet independently verifiable.

At the core of \sys lies a design philosophy that treats incentive execution as a first-class system primitive, while preserving the efficiency of traditional FL workflows. Instead of embedding learning logic directly into blockchain execution, \sys adopts a modular architecture that decouples computation-intensive training and aggregation from economic settlement, enabling scalable and accountable collaboration under open participation.

Specifically, \sys decomposes the FL workflow into four functional modules: \textit{configuration}, \textit{training}, \textit{aggregation}, and \textit{contribution evaluation}. These modules are algorithm-agnostic and can be instantiated with different learning, aggregation, and evaluation strategies, allowing flexible adaptation to diverse application scenarios and incentive policies.

To support for these four functional modules, we design one Web3-based supporting layer with three components: \textit{trusted transactions}, which leverage smart contracts to record configurations, enforce participation rules, and automate incentive settlements; \textit{information communication}, which employs peer-to-peer protocols for efficient model exchange and uses the blockchain only to anchor critical metadata like content identifiers (CIDs), minimizing on-chain overhead; and \textit{information protection}, which integrates cryptographic methods to secure model updates to ensure off-chain evaluations combined with on-chain verification for correctness without exposing raw data.

Together, these modules constitute the operational backbone of \sys. They enable federated learning workflows that are economically accountable, communication-efficient, and secure by design, thereby supporting sustainable collaboration in open Web3 environments.

\subsection{\sys Modular Functional Design}

In \sys, the FL workflow is decomposed into four functional modules: configuration, training, aggregation, and contribution evaluation. These modules are independent in their responsibilities yet interoperable in execution, ensuring that the system can flexibly adapt to different FL algorithms while remaining efficient and reliable in decentralized environments.
This modularity allows researchers or practitioners to extend \sys with novel schemes simply by customizing these modules.

\subsubsection{Configuration module}
The configuration module is responsible for task initialization by the customer. Through a dedicated front-end panel, the customer specifies the essential parameters for an FL project, including model architecture, optimizer type, learning rate, number of iterations, and pricing strategy. In addition, the customer defines the allocation strategy of incentive payments and selects the algorithms for aggregation and contribution evaluation.
This module ensures transparency and reproducibility of task setup, as critical configurations are committed on-chain through smart contracts. Trainers can inspect these parameters before joining, which enhances fairness and accountability.

\subsubsection{Training module}
The training module captures the local training process carried out by trainers. After retrieving the latest global model from the communication layer, each trainer trains it using its private dataset according to the hyperparameters defined in the configuration phase.
To protect information security, local updates are encrypted before leaving the trainer's device. Once encrypted, the updates are uploaded via the communication layer and referenced on-chain. This modular design allows the training logic to remain unchanged while flexibly accommodating diverse FL variants, such as personalized learning, meta-learning, and Byzantine-robust training methods. By aligning with the aggregation module, it further enables seamless integration of algorithm-specific features, adaptive learning rates, or differential privacy mechanisms, ensuring that the system remains extensible and adaptable to emerging FL approaches.

\subsubsection{Aggregation module}
The aggregation module is executed by the aggregator. It collects encrypted model updates from participating trainers, decrypts them, and computes the new global model according to the pre-specified aggregation algorithm. Common strategies such as weighted averaging (FedAvg~\cite{mcmahan2017communication}) or proximal optimization (FedProx~\cite{li2020federatedmlsys}) can be plugged in without altering the rest of the framework. The resulting global model is re-encrypted and distributed to trainers for the next round.
Beyond this core functionality, the aggregation module is designed to be extensible. Rather than embedding specialized defenses or personalization techniques directly into the framework, \sys delegates such capabilities to algorithmic choices within the aggregation process. This separation of concerns ensures that the system architecture remains lightweight, while still supporting the integration of advanced aggregation strategies developed by the FL community.

\subsubsection{Contribution evaluation module}
The contribution evaluation module measures the value of each trainer's update. Once the global model is aggregated, the aggregator evaluates the marginal or relative impact of every submitted update. By decoupling this module, \sys allows users to experiment with different contribution assessment mechanisms while maintaining compatibility with the rest of the workflow.
\sys integrates Shapley value~\cite{shapley1953value} and leave-one-out (LOO)~\cite{wang2020principled} as built-in contribution evaluators. The evaluation module is designed to be pluggable, so new estimators can be added by swapping this module without touching training, aggregation, or settlement logic.

\subsection{Design of the Web3 Supporting Layer}
While the functional modules define the logical workflow of federated learning, the Web3 supporting layer operationalizes this workflow in open environments through three supporting components.
They provide the system mechanisms for transaction management, communication, and information protection, and govern how trust, coordination, and incentives are enforced in practice.
A central design question at this layer is how aggregation and contribution evaluation are executed and certified.

\subsubsection{Architecture and Design Trade-offs}

A fundamental design choice in incentive-aware federated learning is whether aggregation and contribution evaluation are executed in an \emph{owner-executed} mode, where the customer (or a designated trusted aggregator) performs these steps, or in a \emph{committee-executed} mode, where multiple independent parties certify the results via majority agreement. In both cases, we use Web3 primitives for auditability and automated settlement, rather than for executing learning workloads.

A committee-executed architecture improves fault tolerance and reduces reliance on any single executor, but typically incurs high coordination overhead, complex certification logic, and substantial communication costs, especially when evaluation requires access to validation data.
Moreover, distributing evaluation workloads across multiple parties often necessitates exposing shared validation resources, which introduces additional security and privacy risks.

In contrast, an owner-executed architecture enables efficient execution, preserves evaluation data confidentiality, and simplifies system orchestration, while it requires an honest aggregator assumption for correct execution.

In \sys, we prioritize the owner-executed mode as the default deployment setting, as it offers significantly better efficiency and practicality for large-scale, real-world deployments.
Web3 does not replace centralized computation in this mode. Instead, the blockchain acts as an immutable audit ledger and an automated settlement layer: it records round-wise commitments (\eg, aggregated-model references and reward allocations) and executes payouts via smart contracts.
Thus, even with a single executor, incentive outcomes remain transparent and publicly auditable, raising the cost of misconduct and enabling accountability under open participation.
Participants can also exit when unsatisfied, which limits prolonged exposure to unfair execution but does not replace stronger assurance mechanisms.
Unless otherwise specified, the following design describes the owner-executed mode.
At the same time, our architecture is not restricted to a single mode.
The supporting layers are designed to accommodate both owner-executed and committee-executed aggregation, and we discuss the additional mechanisms required for committee execution in \S\ref{sec:decentralized-support}.

\subsubsection{Trusted Transaction}

In \sys, trust among customers, trainers, and aggregators\footnote{When customers take responsibility for aggregation and contribution evaluation, they effectively act as aggregators. Throughout this paper, we restrict our attention to this merged setting and, for brevity, use ``customer'' to represent the combined role.}
is established through smart contracts deployed on blockchain.
The smart contract in \sys serves four key purposes: (1) recording critical project information, (2) orchestrating training progress through event-driven mechanisms, (3) granting rights based on verified payments and roles, and (4) automating incentive distribution. By embedding these rules on-chain, customers, trainers and aggregators can interact transparently, with ownership and access rights automatically enforced.

\noindent\textbf{Recording critical information.} The contract maintains a traceable log of essential project data, including account addresses and public keys of the customer, trainers, and aggregator, as well as fees paid to join the project. For each training iteration, it tracks enrolled participants\footnote{In the remainder of this paper, we use the term ``participants'' to collectively denote customers, aggregators and trainers.} and stores symmetric keys encrypted for them under the dynamic key mechanism.

\noindent\textbf{Coordinating training progress.} To enable seamless interaction, the contract defines four events: \emph{key\_request}, \emph{key\_achieved}, \emph{global\_model\_updated}, and \emph{local\_training\_finished}. Blockchain listeners capture these events, allowing customers, trainers, and aggregators to respond autonomously to state changes. For example, when a trainer joins an iteration, the contract emits \emph{key\_request}, prompting the customer to provide an encrypted key. Once the key is delivered, the \emph{key\_achieved} event allows the trainers to begin local training. Upon completion, \emph{local\_training\_finished} signals the customers to aggregate updates and publish the new global model CID, followed by \emph{global\_model\_updated} for trainers to proceed to the next round.

\noindent\textbf{Granting rights.} Access control is automatically enforced based on on-chain records. Only the customer can update project-critical states, such as global model CIDs, pricing, or continuation decisions. Similarly, only trainers who have paid the required fees are authorized to request decryption keys and access the global model. These guarantees are inherited from the security properties of the underlying blockchain~\cite{wood2014ethereum}.

\noindent\textbf{Automating incentive distribution.} After each training iteration, the customer computes the contribution scores of trainers using the configured evaluation algorithm. The resulting contribution list is submitted to the blockchain as a transaction. Once received, the smart contract verifies the validity of the submission and automatically executes the incentive distribution according to the pre-defined pricing and allocation rules. This mechanism eliminates the need for centralized payment settlement, guarantees that trainers are compensated fairly for their contributions, and ensures that the entire incentive process remains transparent and tamper-proof.

Through this design, the smart contract ensures fairness in transactions and access management while maintaining a transparent and verifiable record of project progress. It establishes the foundation of trusted transactions, upon which communication and protection mechanisms further sustain the secure execution of FL in Web3.

\subsubsection{Information Communication}
\label{sec:datatrans}
To optimize information exchange, model updates, local changes, and non-critical metadata are exchanged off-chain through direct peer-to-peer connections. This approach ensures low-latency communication, making full use of the available network resources for faster and more scalable model updates. By avoiding the overhead of frequent on-chain transactions, we significantly reduce the blockchain's load, keeping on-chain activity focused on critical events, such as cryptographic commitments for model integrity.

To further enhance reliability and fault tolerance, critical data, including model updates, can be optionally anchored on-chain, ensuring verifiability without unnecessary redundancy. For data storage and retrieval, decentralized file systems can be used as a backup to store and share references, providing efficient and fault-tolerant access to model weights across participants.

This hybrid communication model, relying primarily on peer-to-peer communication with optional blockchain anchoring, ensures a balance of efficiency, scalability, and security. It maximizes data transfer speed between clients while maintaining the necessary traceability and integrity of critical updates.

\subsubsection{Information Protection}
\label{sec:cryptosys}

In decentralized systems such as blockchain and IPFS, data transparency ensures verifiability but also risks exposing sensitive information, as semi-honest participants may infer others' updates. To preserve confidentiality during FL, \sys integrates a cryptographic component securing both model uploads and downloads.

\noindent\textbf{ECDH Key Exchange Mechanism.}
While hybrid RSA-AES schemes have been used for secure communication~\cite{khanezaei2014framework,behera2021federated}, Elliptic Curve Diffie-Hellman (ECDH) achieves equivalent security with shorter keys, making it preferable in blockchain environments with limited storage and computation~\cite{gupta2002performance}.

ECDH enables two devices to establish a shared secret without directly exchanging it. Each device generates a key pair $(sk_i, pk_i)$ derived from elliptic curve operations. After exchanging public keys, both parties independently compute the same shared secret $K_{ij} = h(sk_i, pk_j)$, where $h$ denotes the ECDH key derivation function. This shared secret can then be used for encryption and decryption of transmitted data.

In \sys, the customer generates $(sk_0, pk_0)$, while each trainer $i$ generates $(sk_i, pk_i)$. Public keys are registered on the blockchain, enabling both sides to compute their shared key $K_i$:
\begin{equation}
    K_i = h(sk_0, pk_i) = h(sk_i, pk_0).
\end{equation}

Trainers use $K_i$ to encrypt their local model updates before uploading them to IPFS (or transmitting via the primary communication channel). The customer retrieves and decrypts these updates with the same key, ensuring confidentiality and correctness throughout the upload process.

\noindent\textbf{Dynamic Symmetric Key Mechanism.}
While ECDH is effective for securing individual uploads, using it directly for global model distribution is inefficient, as the customer would need to encrypt the model separately for each trainer. To address this, \sys employs a dynamic symmetric key mechanism during the download phase.

At the start of each iteration, the customer generates a temporary symmetric key and distributes it securely by encrypting it under each trainer's ECDH-derived key. All trainers then share the same symmetric key for decrypting the global model in that round. This design significantly reduces time and storage overhead while maintaining the same level of confidentiality as direct ECDH encryption.

To further strengthen security, the symmetric key can be refreshed in every iteration. This ensures fine-grained access control, limits the risk of key exposure, and guarantees that only legitimate trainers can access the global model during its validity period.

\subsubsection{Supporting Committee-Executed Mode}
\label{sec:decentralized-support}

While \sys prioritizes the \emph{owner-executed} mode by default, the supporting layers also accommodate a \emph{committee-executed} mode when deployments require reduced reliance on a single executor.
Enabling committee execution only requires localized changes to the three supporting layers: the contract records and certifies majority-endorsed results, the communication layer routes downloads to certified aggregators, and the protection layer remains unchanged.

\noindent\textbf{Trusted transaction.}
In the committee-executed mode, multiple aggregators independently perform off-chain aggregation and contribution evaluation.
Each aggregator submits the resulting model CID and contribution list to the contract. The contract adopts the version endorsed by a majority of aggregators and records the endorsers as the certified providers for the current round.

\noindent\textbf{Information communication.}
Trainers obtain the next-round global model only from certified providers. Concretely, a trainer queries the on-chain endorser list and sequentially requests the model from certified aggregators until a valid response is received.

\noindent\textbf{Information protection.}
The protection mechanism remains unchanged across modes.
Global models are encrypted with a symmetric key generated by the customer and distributed via ECDH. Local updates are encrypted by trainers and shared only with aggregators, ensuring confidentiality regardless of how aggregation and evaluation are executed.

Overall, this design restricts on-chain operations to lightweight majority certification and state commitment, while keeping computation and data transmission off-chain. As a result, committee-executed aggregation can be supported without incurring prohibitive on-chain storage or execution overhead.

%% file: sec/4_impl.tex
\section{Implementation}

This section describes how the system is realized in a web-native environment. Our implementation emphasizes a single platform that simultaneously serves end users (who train directly in the browser) and researchers (who orchestrate large-scale experiments and prototype new aggregation or reward schemes).



\subsection{User- and Developer-Friendly Platform}
\label{sec:impl:dual}

\subsubsection{Web-Native Client Runtime}
\label{sec:impl:runtime}

The client runtime of \sys is implemented in TypeScript and runs entirely in commodity browsers.
We use \texttt{TensorFlow.js}, selecting WebGL when GPU support is available and falling back to WASM for CPU-only environments.
Numerical tensors are stored as typed arrays with reuse to mitigate garbage-collection overhead.
Datasets and intermediate artifacts are persisted in \emph{IndexedDB}, while lightweight client states are maintained in \emph{LocalStorage}.
This separation reduces serialization overhead and enables predictable recovery after tab suspension or crashes.

\subsubsection{Experiment Control and Cohort Management}
\label{sec:impl:control}

To support controlled experimentation and large-scale evaluation, \sys provides a cohort management layer for automated client orchestration.
The control service can launch and configure batches of browser clients, handle cryptographic key generation, fund blockchain accounts, and perform setup tasks required for reproducible experiments.
A web-based interface supports bulk configuration and remote control, while telemetry modules collect logs and performance metrics across clients in a unified format.
This infrastructure enables systematic evaluation under consistent deployment conditions.

\subsubsection{Developer Extensibility and Tooling}
To facilitate algorithmic experimentation, \sys exposes stable plug-in interfaces for aggregation and reward functions.
Researchers can integrate alternative aggregation rules, contribution evaluation methods, or incentive policies without modifying the core runtime, communication layer, or smart-contract logic.

Beyond algorithm interfaces, \sys provides an experimentation harness for large-scale and reproducible runs.
In particular, we implement a headless batch browser launcher that can spawn and manage cohorts of Web clients with consistent configurations, enabling controlled ablations and stress tests under realistic network conditions.
The accompanying web UI supports bulk configuration, remote control, and telemetry collection (logs and key metrics) across the cohort.
It also exposes smart-contract states for real-time inspection, simplifying debugging and improving operational visibility.

\subsection{Low-Latency Update Dissemination}
\label{sec:impl:peerprotocol}

We implement the hybrid communication design of \sys using a combination of WebRTC-based peer-to-peer channels and decentralized storage services.
Model updates, control messages, and non-critical metadata are primarily exchanged through direct WebRTC connections, enabling low-latency transmission and efficient utilization of available network bandwidth.

To ensure robustness under unstable network conditions, we integrate IPFS as a secondary communication and storage layer.
When direct peer-to-peer delivery fails, model updates can be retrieved from IPFS using content identifiers recorded on-chain.
This fallback mechanism improves availability and fault tolerance without introducing persistent dependence on distributed storage.

For deployment at scale, we operate a dedicated signaling server to coordinate WebRTC session establishment among large numbers of participants.
In environments where direct peer connectivity is restricted by NATs or firewalls, a TURN (Traversal Using Relays around NAT) server is used to relay traffic and maintain connectivity.
These auxiliary services are lightweight and only assist connection setup and fallback delivery; they do not participate in model processing or incentive execution.

Overall, this implementation realizes the hybrid communication design of \sys in practice.
WebRTC serves as the primary low-latency data plane, IPFS provides resilient backup storage, and the blockchain anchors integrity-critical references.
Together, they enable efficient, reliable, and verifiable model exchange in heterogeneous network environments.

%% file: sec/5_exp.tex
\section{Experiments}
The experiments are designed to achieve three primary goals:

\noindent\textbf{System Performance Evaluation}. 
We aim to demonstrate that \sys's Web3 integration, particularly the blockchain-based smart contract execution, introduces reasonable non-training overhead, with the execution time and gas fees remaining within acceptable limits.

\noindent\textbf{Conformance Experiments}. To validate \sys's ability to train FL models, ensuring the correctness and feasibility of the training process, and assessing model accuracy as a key indicator of system performance.

\noindent\textbf{Usability and Extensibility}. 
We evaluate \sys's practicality by measuring deployment and onboarding effort against representative blockchain-based FL frameworks, as well as the development effort required to integrate classic aggregation and contribution-evaluation algorithms.

\subsection{System Demonstration}

\begin{figure}[t]
    \centering
    \subfigure[User control panel of \sys]{\includegraphics[width=0.95\linewidth]{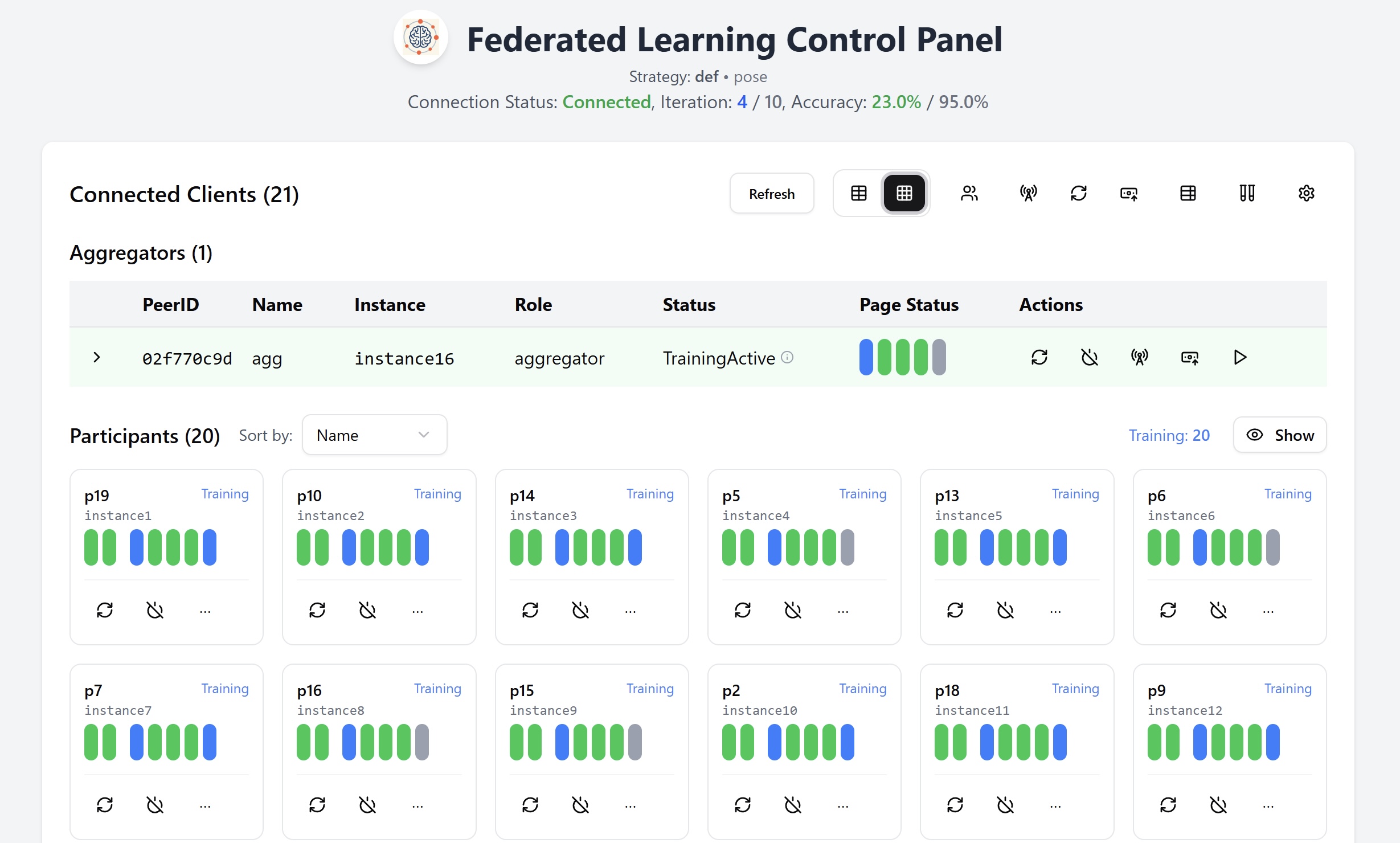}\label{fig:demo-1}}\\
    \subfigure[Log information and collected metrics]{\includegraphics[width=0.95\linewidth]{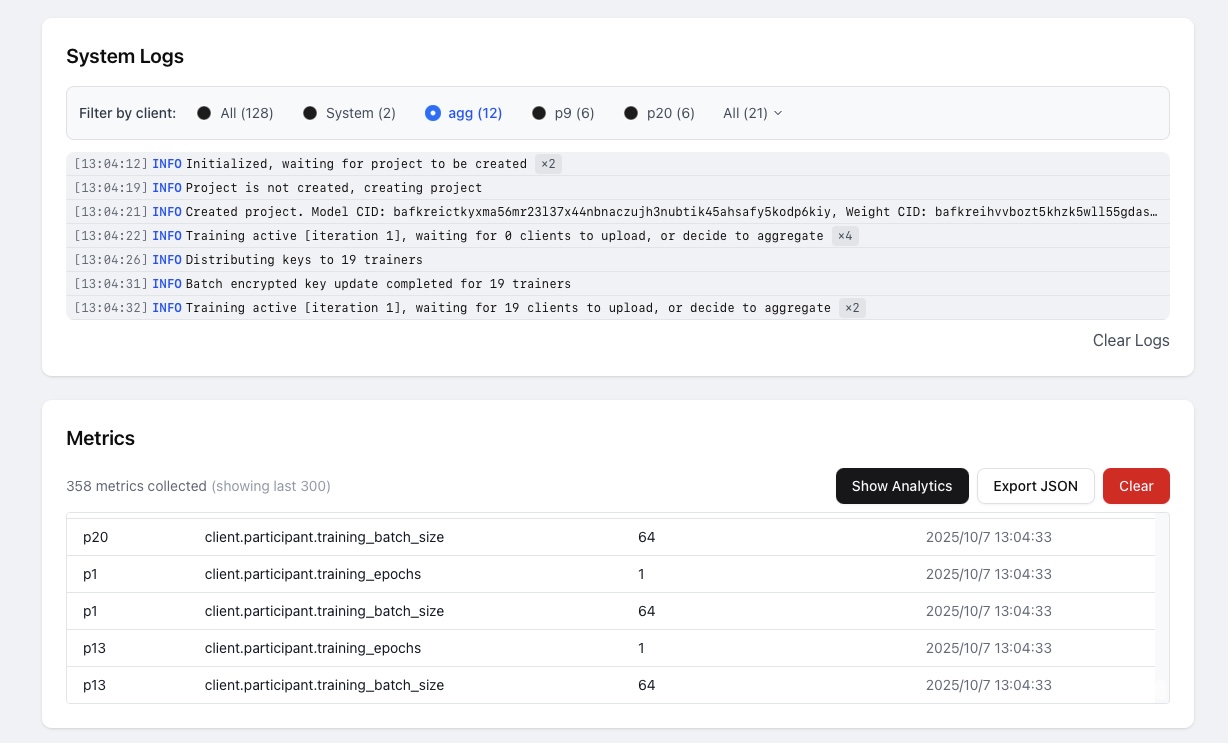}\label{fig:demo-2}}
    \caption{Demonstration of \sys. The user control panel shows the status of all aggregators and trainers, including peer connections, IPFS availability, dataset readiness, and training progress. System logs and key metrics are also displayed for monitoring and debugging purposes.}
    \label{fig:ui-demo}
\end{figure}

To showcase the usability and practical deployment of \sys, we provide a demonstration of the system's user interface. The UI allows users to configure FL tasks, monitor the training process, and observe aggregation and contribution evaluation results in real time.
As shown in Figure~\ref{fig:ui-demo}, the interface clearly displays task progress, model performance metrics, and device status, providing an intuitive overview of the workflow. This demonstration highlights that \sys is not only functionally complete but also accessible to users without deep technical expertise, reinforcing its practicality for real-world FL scenarios.

\begin{figure*}[t]
    \centering
    \subfigure[Block Generation Time = 5s, network latency = 50ms]{\includegraphics[width=0.8\linewidth]{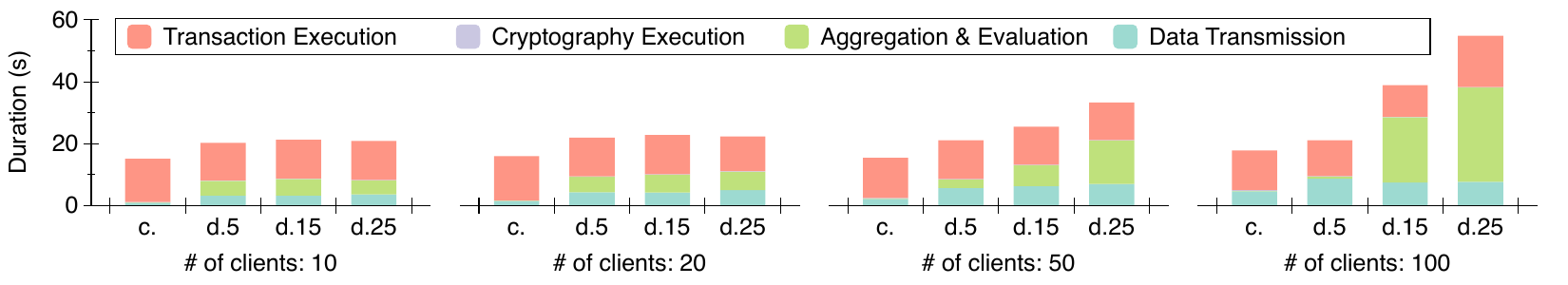}\label{fig:statis-5s}}\\
    \vspace{-0.1in}
    \subfigure[Block Generation Time = 12s, network latency = 50ms]{\includegraphics[width=0.8\linewidth]{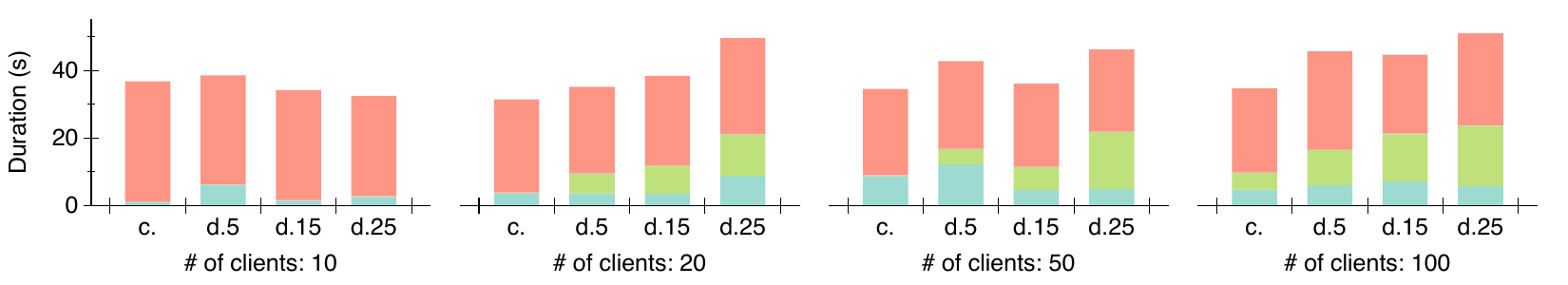}\label{fig:statis-12s}}\\
    \vspace{-0.1in}
    \subfigure[{Block Generation Time = 5s, network latency = 200ms}]{\includegraphics[width=0.8\linewidth]{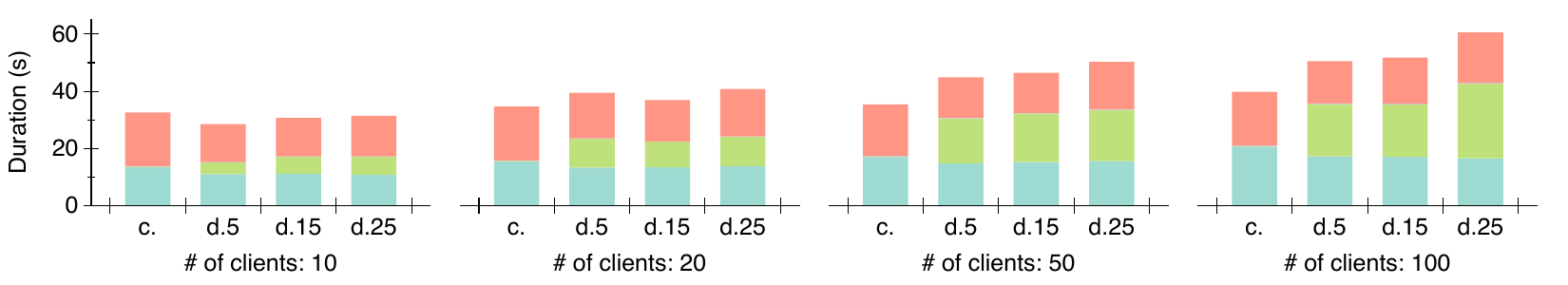}\label{fig:statis-200ms-5s}}\\
    \vspace{-0.1in}
    \subfigure[{Block Generation Time = 12s, network latency = 200ms}]{\includegraphics[width=0.8\linewidth]{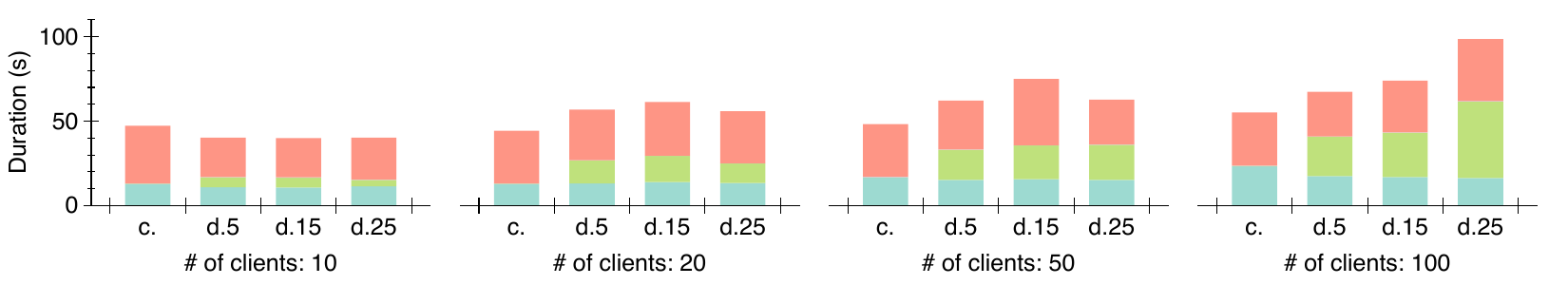}\label{fig:statis-200ms-12s}}\\
    \caption{{Per-round non-training time overhead under owner-executed (single aggregator) and committee-executed modes (committee size 5/15/25), with 10/20/50/100 trainers.
We compare two block generation intervals (5\,s vs.\ 12\,s) and two network latency (50\,ms vs. 200\,ms). Larger committees increase coordination overhead.} Here, \texttt{c.} denotes the owner-executed mode, while \texttt{d.5}, \texttt{d.15}, and \texttt{d.25} denote committee-executed modes with committee sizes of 5, 15, and 25, respectively.}
    \label{fig:aggregators-statistics}
\end{figure*}

\subsection{Feasibility Comparison of Owner-Executed and Committee-Executed Modes}
\label{sec:feasibility-modes}

We evaluate the feasibility of \sys under different execution modes by comparing the owner-executed setting (single aggregator) with the committee-executed setting (multiple aggregators jointly endorsing aggregation and evaluation results).
Figure~\ref{fig:aggregators-statistics} reports the per-round \emph{non-training} time overhead when varying the number of trainers (10/20/50/100) and the committee size (5/15/25), under two block generation intervals (5\,s and 12\,s).

To emulate realistic wide-area deployments, we further control network conditions using Linux traffic control (\texttt{tc}).
In addition to a baseline WAN setting with 100\,Mbps bandwidth and 50\,ms round-trip latency, we introduce a high-latency configuration with 200\,ms delay to approximate cross-continent communication.
These settings allow us to assess the robustness of different execution modes under heterogeneous network environments.

\subsubsection{Under different block generation time.}
Figure~\ref{fig:statis-5s} and Figure~\ref{fig:statis-12s} illustrate the per-round non-training overhead under different block generation intervals.
Overall, both execution modes remain operational across all tested configurations, indicating that \sys can reliably complete training rounds without interfering with learning computation.

We observe that the committee-executed mode consistently incurs higher overhead than the owner-executed mode, and this gap widens as the committee size increases due to additional coordination, cross-verification, and endorsement steps.

Block generation time further shapes this overhead: increasing the interval from 5\,s to 12\,s enlarges the confirmation window and makes transaction execution the dominant contributor, which amplifies coordination-induced waiting and results in noticeably higher round latency, especially in committee-based settings.
A shorter interval (5\,s) partially amortizes transaction latency and mitigates this effect, but committee execution still shows a persistent overhead premium over the owner-executed baseline.

\subsubsection{Under different network latency.}

Figure~\ref{fig:statis-200ms-5s} and Figure~\ref{fig:statis-200ms-12s} further report the impact of network latency on per-round non-training overhead.
Compared with the baseline WAN setting, increasing round-trip latency from 50\,ms to 200\,ms leads to a consistent growth in coordination and synchronization time across all execution modes.

Even under high-latency conditions, the additional overhead remains moderate and grows slower than the increase in the number of participants. 
Since model dissemination and aggregation proceed through optimized off-chain channels, while endorsement and settlement are decoupled from the training path, increased network latency mainly affects peer-to-peer communication and does not compound with block confirmation delays.


Taken together, these results suggest that while both modes are feasible, the owner-executed mode is substantially more time-efficient and more robust to blockchain latency, making it a preferred default for large-scale deployments where efficiency is the primary concern.

\subsection{System Efficiency under Web3 Constraints}

\subsubsection{Experimental Setup}
To evaluate the performance of our Web3-based FL framework under different network condition, we conduct experiments under two distinct network configurations:

\noindent\textbf{LAN topology.} In this setup, the contract is deployed on a local test chain launched using Foundry\footnote{https://github.com/foundry-rs/foundry}. We use 6 Mac devices with M-series chips, one acting as the customer and the other five as trainers. All devices are connected within a local network with 100 Mbps bandwidth, ensuring minimal network and transaction latency. The application services are also deployed within this LAN.

\noindent\textbf{WAN topology.} In this setup, the contract is deployed on Sepolia (Ethereum testnet). We lease 21 devices, each equipped with an Intel Core i5-14600KF CPU or a comparable processor, from a cloud service provider\footnote{https://vast.ai} with devices distributed across multiple regions worldwide. One device serves as the customer, and the remaining 20 devices act as trainers. This configuration reflects the challenges and performance characteristics encountered in actual deployment scenarios, where devices are distributed across multiple regions.

The LAN configuration simulates a controlled development environment, while the WAN setup represents a geographically dispersed scenario typical of real-world deployments.

For the experiments, we adopt ResNet-18~\cite{he2016deep} as the global model and partition the CIFAR-10~\cite{krizhevsky2009learning} dataset into $n$ non-overlapping IID subsets, each assigned to a different trainer. Training is executed for 100 rounds, and the results are reported as averages to account for variability. We measure key metrics, including phase-wise overhead and the associated gas fees in both LAN and WAN environments.

\subsubsection{Results and Analysis}

\input{tabs/time}

Table~\ref{tab:execution_time} reports the breakdown of execution time across different phases under LAN and WAN environments.
In both settings, local training remains the dominant component of total runtime, accounting for 54.2\% under LAN and 55.8\% under WAN, indicating that model computation continues to be the primary performance bottleneck.

When transitioning from LAN to WAN, the overall round time increases from 33.74\,s to 152.81\,s.
This growth is mainly driven by the aggregation and evaluation phase, whose absolute runtime rises from 1.96\,s to 28.48\,s, and by the prolonged training phase under heterogeneous deployment conditions.
In particular, the WAN setup relies on CPU-only cloud instances, whereas the LAN environment benefits from M-series devices with stronger on-device acceleration.

Transaction execution constitutes the second largest overhead under the LAN setting (35.9\%), primarily because the local test chain uses a shorter block generation interval and the overall computation workload is relatively small.
Under the WAN setting, the absolute transaction time increases to 32.57\,s.
This increase is consistent with the slower block generation rate on the public testnet (Sepolia), as the magnitude of the additional delay aligns with the longer per-block interval compared to the local test chain.

Cryptographic operations, including encryption, decryption, and key generation, account for less than 1\% of total runtime in both environments, demonstrating that security mechanisms are implemented efficiently with negligible performance impact.
Similarly, data transmission remains a minor component, contributing only 3.4\% of total runtime under both LAN and WAN settings.

We further evaluate the economic cost of blockchain interaction in \sys.
Each on-chain operation, including task creation, commitment submission, aggregation finalization, and reward distribution, incurs a modest gas fee.
The average gas consumption per round is 1,249,407 for LAN and 3,506,309 for WAN.
Using the 7-day average gas price\footnote{\url{https://etherscan.io/gastracker}} of 0.084\,gwei (Feb.~21--27,~2026) and an ETH/USD exchange rate\footnote{\url{https://etherscan.io/chart/etherprice}} of 1950.46, the corresponding monetary cost per round is approximately \$0.205 for LAN and \$0.575 for WAN.

Overall, these results demonstrate that Web3 integration in \sys introduces bounded computational and monetary overhead while preserving the efficiency of standard FL workflows.
As training and aggregation dominate system cost under realistic deployments, blockchain and cryptographic components remain secondary factors, enabling practical and scalable operation in decentralized environments.

\subsubsection{Training Conformance}
After 100 rounds of FL training and aggregation, the global model achieves an accuracy of 61.44\% in the LAN setting (5 clients) and 55.39\% in the WAN setting (20 clients), demonstrating that our system maintains reasonable model performance under different network configurations.
This is as expected because our framework does not change any setting of the training and aggregation process of the traditional FL. The data transmission is also lossless.

\subsection{Usability and Extensibility}
Deploying federated learning frameworks integrated with blockchain infrastructures is often non-trivial, as such systems involve multiple primitives, including smart contracts, cryptographic protocols, decentralized storage, and distributed training runtimes. These dependencies significantly increase the operational burden and hinder practical adoption.

\input{tabs/usability}
To evaluate the usability of FWeb3, we compare its deployment effort with two open-source blockchain-based FL frameworks, namely VeryFL~\cite{li2025veryfl} and OpenFL~\cite{wahrstatter2024openfl}. We measure both operational complexity using objective indicators and temporal cost, including manual configuration overhead and deployment time, as reported in Table~\ref{tab:deployment} and Table~\ref{tab:deploy_time}. 

As shown in Table~\ref{tab:deployment}, although VeryFL and OpenFL are mainly evaluated under local experimental settings, FWeb3 requires fewer manual steps, runtime components, executed commands, and configuration files, indicating substantially lower deployment complexity. 
Moreover, FWeb3 further supports deployment on public blockchains and over wide-area networks with minimal additional configuration.
Detailed deployment procedures are provided in Appendix~\ref{app:deployment}.

We measure deployment usability with three complementary metrics that capture both the platform-operator and participant perspectives. 
\textbf{Time-to-First-Run (TFR)} captures the operator time from repository cloning to setting up a \emph{small-scale experimental} environment ready for a minimal end-to-end FL run by following the official documentation. 
In contrast, \textbf{Time-to-Usable Deployment (TUD)} measures the operator time to bring the framework to a \emph{practically deployable, business-usable} state, where it can reliably support multi-round training with incentive settlement enabled and provides basic operational capabilities such as task configuration. 
Finally, \textbf{User Onboarding Time (UOT)} measures the participant time for a new user with zero prior experience to complete the required setup and successfully join an existing training task, and start one training round as a participant.

As shown in Table~\ref{tab:deploy_time}, for VeryFL and OpenFL, only TFR is reported, as its current implementation mainly supports local-chain simulation and does not support inter-client communication via internet and real-world deployment.
As a result, TUD and UOT are not practically applicable.
Even under this limited setting, achieving a runnable environment requires around five minutes and involves non-trivial manual configuration.
In contrast, FWeb3 achieves consistently low overhead across all metrics.
Operators only need to install lightweight dependencies and configure basic wallet and network parameters, and both local and public-chain deployments follow unified workflows, leading to TFR and TUD below three minutes.
From the participant perspective, users can join training tasks directly through the browser-based interface, resulting in UOT below one minute.

Overall, these results demonstrate that FWeb3 substantially reduces deployment and onboarding overhead, enabling rapid setup and frictionless participation in practical environments.

\input{tabs/extensibility}

Beyond deployment usability, we further assess the extensibility of FWeb3 in supporting algorithmic evolution. 
We extend FWeb3 along two core extension points: \emph{aggregation} and \emph{contribution evaluation}. 
Specifically, we implement two additional aggregation algorithms and two classical contribution evaluation methods, and quantify the engineering effort using the number of modified/new source files and the lines of code changed (LOC). 
As summarized in Table~\ref{tab:extensibility}, integrating a new aggregator requires modifying only 3 files with fewer than 60 added LOC, while adding a new contribution evaluator requires only 2 files with around 80 added LOC. 
Overall, the consistently small code footprint across all extensions indicates that FWeb3 cleanly decouples Web3-specific components from algorithmic logic, enabling lightweight integration of new methods without re-engineering the underlying system pipeline.

%% file: tabs/time.tex
\begin{table}[t]
\centering
\caption{{Breakdown of execution time across different phases under LAN and WAN environments.}}
\label{tab:execution_time}
\setlength{\tabcolsep}{1mm}
\begin{tabular}{lcccc}
\toprule
\multirow{2}{*}{Phase} 
& \multicolumn{2}{c}{LAN} 
& \multicolumn{2}{c}{WAN} \\
\cmidrule(lr){2-3} \cmidrule(lr){4-5}
& Time (s) & (\%) & Time (s) & (\%) \\
\midrule
Local Training            & 18.28 & 54.2 & 85.29 & 55.8 \\
Aggregation and Evaluation& 1.96  & 5.8  & 28.48 & 18.6 \\
Cryptography Execution    & 0.24  & 0.7  & 1.31  & 0.9 \\
Transaction Execution    & 12.12 & 35.9 & 32.57 & 21.3 \\
Data Transmission         & 1.14  & 3.4  & 5.16  & 3.4 \\
\midrule
Total                     & 33.74 & 100  & 152.81 & 100 \\
\bottomrule
\end{tabular}
\end{table}

%% file: tabs/usability.tex
\begin{table}[t]
\centering
\caption{Deployment Effort Comparison.}
\label{tab:deployment}
\setlength{\tabcolsep}{1mm}
\begin{tabular}{lcccc}
\toprule
\textbf{Framework} 
& \#Steps 
& \#Components 
& \#Commands 
& \#Configs \\
\midrule
VeryFL~\cite{li2025veryfl} & 9 & 4 & 4 & 3 \\
OpenFL~\cite{wahrstatter2024openfl} & 7 & 5 & 6 & 3 \\
\textbf{FWeb3 (Ours)}        & 6  & 4 & 2  & 1 \\
\bottomrule
\end{tabular}
\end{table}

\begin{table}[t]
\centering
\caption{Deployment and Onboarding Time Comparison.}
\label{tab:deploy_time}
\begin{tabular}{lccc}
\toprule
\textbf{Framework} & \textbf{TFR} & \textbf{TUD} & \textbf{UOT} \\
\midrule
VeryFL~\cite{li2025veryfl}       & $\sim$5min & \textcolor{red}{---} & \textcolor{red}{---} \\
OpenFL~\cite{wahrstatter2024openfl} & $\sim$10min & \textcolor{red}{---} & \textcolor{red}{---} \\
\textbf{FWeb3 (Ours)} & <3min & <3min & <1min \\
\bottomrule
\end{tabular}

\end{table}

%% file: tabs/extensibility.tex
\begin{table}[t]
\centering
\caption{Extensibility Evaluation of FWeb3. Files denotes the number of modified or newly added source files. LOC denotes lines of code changed.} %
\label{tab:extensibility}
\begin{tabular}{llcc}
\toprule
\textbf{Extension Task} 
& \textbf{Algorithm} 
& \textbf{Files} 
& \textbf{LOC} \\
\midrule
\multirow{2}{*}{Aggregation} &  FedAvgM~\cite{hsu2019measuring}  & 3 & \textcolor{darkgreen}{+48} \textcolor{red}{-3} \\
& FedAdam~\cite{reddi2021adaptive}  & 3 & \textcolor{darkgreen}{+56} \textcolor{red}{-2} \\ \midrule
\multirow{2}{*}{\makecell{Contribution\\Evaluation}} & CGSV~\cite{xu2021gradient} & 2 & \textcolor{darkgreen}{+79} \textcolor{red}{-0} \\
&  GTG-Shapley~\cite{liu2022gtg} & 2 & \textcolor{darkgreen}{+77} \textcolor{red}{-4} \\
\bottomrule
\end{tabular}

\end{table}

%% file: sec/6_conclusion.tex
\section{Conclusion}

In this paper, we present \sys, a practical Web3-enabled federated learning framework for incentive-aware learning in open-participation settings. \sys treats incentive execution as a first-class systems concern and organizes the workflow through a modular design that separates federated learning functions from Web3 support services. This design decouples computation-intensive training and aggregation from blockchain interaction, while preserving traceable settlement and supporting pluggable aggregation and contribution evaluation methods.
We implement \sys in a web-native environment with a browser-based DApp interface for task configuration, participation, and monitoring. Experimental results in LAN and WAN settings show that \sys enables end-to-end incentive-aware federated learning with acceptable non-training overhead and supports practical deployment and user onboarding. Our extensibility study further suggests that new aggregation and contribution evaluation methods can be integrated with a small code footprint.
Overall, \sys provides a reusable systems substrate for bridging incentive-aware federated learning research and practical Web3-based deployment. An important direction for future work is to strengthen execution assurance under weaker trust assumptions, including more verifiable contribution evaluation and broader support for adversarial settings.

%% file: sec/appendix.tex
\appendix
\section{Detailed Deployment Effort Measurement}
\label{app:deployment}

This appendix provides a detailed explanation of how the deployment effort metrics in Table~\ref{tab:deployment} are obtained.
For each framework, we strictly follow the official documentation and repository instructions to deploy a minimal end-to-end federated learning task.
All manual operations, executed commands, required runtime components, cryptographic materials, and configuration modifications are recorded during the deployment process.

A manual step is counted when explicit human intervention is required.
A component is counted when an independent runtime service must be deployed and maintained.
A secret is counted when cryptographic material must be manually generated, stored, or configured.
A command is counted when a shell command must be manually executed.
A configuration is counted when a file must be manually edited.
All measurements are conducted under identical hardware and network settings and repeated three times.

\subsection{VeryFL}

Deploying VeryFL~\cite{li2025veryfl} involves coordinating multiple heterogeneous toolchains and runtime environments, including blockchain emulation, smart contract management, and Python-based training frameworks\footnote{\url{https://github.com/GTMLLab/VeryFL}}. 
As a result, users must perform a sequence of manual operations across different software stacks, leading to non-trivial deployment effort.

\textbf{Manual Steps.}
The deployment process requires a series of explicit human interventions, covering environment preparation, blockchain setup, and training initialization. 
In particular, the main steps include:

\begin{enumerate}
  \item Preparing Node.js environment.
  \item Installing the local blockchain emulator Ganache.
  \item Initializing a Python environment and installing dependencies.
  \item Installing and configuring the python-based development and testing framework for smart contracts targeting EVM, \ie, eth-brownie.
  \item Creating and importing blockchain accounts.
  \item Starting the local Ethereum network.
  \item Compiling and deploying smart contracts.
  \item Configuring benchmark parameters.
  \item Launching training scripts.
\end{enumerate}

These operations must be executed sequentially and often require manual verification, resulting in more than ten deployment steps.

\textbf{Runtime Components.}
From a system perspective, VeryFL relies on several independent runtime components that must be deployed and maintained separately, including:

\begin{itemize}
  \item A local Ethereum emulator (Ganache),
  \item The Brownie contract management framework,
  \item The Python training runtime with PyTorch,
  \item Dataset management utilities.
\end{itemize}

Each component has its own dependency chain and startup procedure, increasing operational complexity.

\textbf{Executed Commands.}
The above setup requires executing several shell commands for dependency installation, service initialization, and task orchestration. 
Representative examples include:

\begin{verbatim}
npm install ganache --global
conda create -n veryfl python==3.9
pip install eth-brownie torch==1.13.0
python test.py --benchmark FashionMNIST
\end{verbatim}

By executing the above commands, VeryFL will launch the simulation on the local blockchain emulator instead of deploying on the public blockchain and internet.

\textbf{Configuration Files.}
Several configuration files must be manually edited to align blockchain settings, training parameters, and benchmark specifications, such as:

\begin{itemize}
  \item benchmark configuration files,
  \item blockchain network settings,
  \item Brownie project configurations.
\end{itemize}

These files must be modified consistently to ensure correct interaction among system components.

Overall, due to its intrinsic design limitations, VeryFL primarily supports simulation on local blockchain emulators and relies on centralized script-based client emulation rather than explicit abstractions for independent participant clients. The need for multi-stage manual operations, multiple runtime components, frequent command execution, and manual configuration contributes to the relatively high deployment effort reported in Table~\ref{tab:deployment}.

\subsection{OpenFL}

OpenFL~\cite{wahrstatter2024openfl} is implemented as a research-oriented decentralized federated learning prototype built on Ethereum smart contracts\footnote{\url{https://github.com/nerolation/OpenFL}}.
The system relies on Jupyter Notebook-based orchestration and manual blockchain configuration, resulting in a multi-stage and non-automated deployment workflow.

\textbf{Manual Steps.}
Deploying OpenFL requires explicit preparation of Python environments, smart contract compilation, blockchain configuration, and notebook execution.
The main steps include:

\begin{enumerate}
  \item Preparing Node.js environment.
  \item Creating a Python virtual environment and installing required python dependencies.
  \item Installing and configuring the local blockchain emulator Ganache.
  \item Preparing configuration files.
  \item Starting the local Ethereum network.
  \item Compiling and deploying smart contracts.
  \item Launching training via Jupyter Notebook.
\end{enumerate}

These steps require manual execution and validation, resulting in a relatively high number of deployment steps.

\textbf{Runtime Components.}
OpenFL depends on several independently managed runtime components, including:

\begin{itemize}
  \item A local blockchain node,
  \item Toolchain for smart contract management,
  \item The Python training runtime with PyTorch,
\end{itemize}

These components are not orchestrated through unified scripts and must be started and monitored manually.

\textbf{Executed Commands.}
The deployment process requires multiple explicit commands\footnote{The original OpenFL repository does not provide unified scripts for contract compilation and deployment, local blockchain initialization, or notebook execution. To ensure reproducibility and standardized measurement, we implemented lightweight automation scripts and released them at \url{https://github.com/KoalaYan/OpenFL}. These scripts encapsulate the above procedures and correspond to the last three commands listed below.} for environment setup, blockchain initialization, and experiment execution, including:

\begin{verbatim}
npm install ganache --global
conda create -n openfl python==3.11
pip install -r requirements.txt
./scripts/build_contracts.sh
./scripts/start_anvil.sh
./scripts/run_notebook.sh
\end{verbatim}

Additional manual edits to configuration files or Python scripts may be required depending on fork mode and chain configuration.

\textbf{Configuration Files.}
OpenFL requires manual modification of several configuration artifacts, including:

\begin{itemize}
  \item blockchain network settings,
  \item blockchain address configuration,
  \item FL training configuration.
\end{itemize}

These configuration steps directly affect blockchain connectivity and transaction construction, and improper configuration may lead to runtime failures.

Overall, the reliance on notebook-driven orchestration, centralized script-based client emulation, and manual blockchain initialization results in substantial deployment effort and limits the system to controlled experimental settings without explicit support for independent participant clients.

\subsection{FWeb3 (Ours)}
To ensure fair comparison with existing blockchain-based federated learning frameworks, all deployment and usability evaluations in this section are conducted under a unified local simulation setting, where a local blockchain emulator is used and multiple client instances are launched on the same machine or local network.
This setting is consistent with the experimental configurations adopted in VeryFL and OpenFL. Notably, although running in a local environment, FWeb3 still deploys multiple independent client processes and enables real peer-to-peer communication among them.

FWeb3 is designed with deployment usability and automation as core objectives.
As reflected in the repository structure and deployment scripts, the system integrates local test blockchain services, WebRTC communication, decentralized storage, and in-browser calculation into a unified orchestration framework.
Most infrastructure components are encapsulated behind standardized Makefile commands and Docker-based workflows, significantly reducing manual engineering overhead.

\textbf{Manual Steps.}
Deploying FWeb3 primarily involves environment preparation and executing high-level orchestration commands.
For local development, users only need to copy environment templates, initialize submodules, and launch the system:

\begin{enumerate}
  \item Preparing Node.js and installing dependencies.
  \item Launching the unified development command.
  \item Configuring participants' role (aggregator or trainer).
  \item Starting a training task via the web interface.
\end{enumerate}

All infrastructure setup is abstracted into a small number of Makefile commands, which automatically prepare the blockchain, deploy smart contracts, initialize backend services, and launch client instances. Compared to prior frameworks, no multi-language compilation or independent service orchestration is required, resulting in substantially fewer manual steps.

\textbf{Runtime Components.}
FWeb3 consolidates system components into a small set of coordinated services, including:

\begin{itemize}
  \item Smart contracts managed via Foundry with a local test blockchain,
  \item A unified backend server handling WebRTC signaling and React-based web services,
  \item An in-browser TensorFlow.js training runtime.
\end{itemize}

These components are launched through centralized orchestration scripts, avoiding fragmented service management across different toolchains.

\textbf{Executed Commands.}
Deployment and runtime management are driven by a limited set of high-level commands, including:

\begin{verbatim}
make build
make up N=$clients
\end{verbatim}

These commands encapsulate contract deployment, blockchain initialization, server startup, and client instantiation, abstracting away low-level configuration details.

\textbf{Configuration Files.}
Most system parameters are specified through standardized environment files (e.g., \texttt{.env}) and unified configuration interfaces.
Users only need to edit a minimal set of fields, such as blockchain provider URLs, private keys, and contract addresses, when switching between different deployment setups.
No manual modification of internal service code or distributed configuration files is required.

Overall, by integrating contract deployment, blockchain initialization, communication setup, and federated training orchestration into a cohesive workflow under a unified local simulation setting, FWeb3 minimizes manual intervention and operational complexity.
This design directly leads to the low deployment effort reported in Table~\ref{tab:deployment}.

\textbf{Public Blockchain and Internet Deployment.}
Beyond local simulation, FWeb3 is designed to support deployment on public blockchains and over wide-area networks.
To migrate from the local setting to real-world deployment, users only need to update environment configuration files (e.g., \texttt{.env}) to specify public RPC endpoints, wallet credentials, and storage services, and redeploy smart contracts using:
\begin{verbatim}
yarn deploy:prod
\end{verbatim}
Core services can then be launched via:
\begin{verbatim}
make build
make core-up
\end{verbatim}
After that, participants can join training tasks through standard web interfaces.
No modification to the system architecture or training logic is required.
This design enables seamless transition from controlled experimental environments to practical large-scale deployments.

%% file: main.bib
@String(AAAI = {AAAI})

@article{meurisch2021data,
  title={Data protection in AI services: A survey},
  author={Meurisch, Christian and M{\"u}hlh{\"a}user, Max},
  journal={ACM Computing Surveys (CSUR)},
  volume={54},
  number={2},
  pages={1--38},
  year={2021},
  publisher={ACM New York, NY, USA}
}

@article{cao2022ai,
  title={Ai in finance: challenges, techniques, and opportunities},
  author={Cao, Longbing},
  journal={ACM Computing Surveys (CSUR)},
  volume={55},
  number={3},
  pages={1--38},
  year={2022},
  publisher={ACM New York, NY}
}

@article{liu2021fate,
  title={{FATE: An Industrial Grade Platform for Collaborative Learning With Data Protection}},
  author={Liu, Yang and Fan, Tao and Chen, Tianjian and Xu, Qian and Yang, Qiang},
  journal={Journal of Machine Learning Research},
  year={2021}
}

@inproceedings{oldenhof2023industry,
  author={Oldenhof, Martijn and {\'A}cs, Gergely and Pej{\'o}, Bal{\'a}zs and Schuffenhauer, Ansgar and Holway, Nicholas and Sturm, No{\'e} and Dieckmann, Arne and Fortmeier, Oliver and Boniface, Eric and Mayer, Cl{\'e}ment and others},
  title={{Industry-Scale Orchestrated Federated Learning for Drug Discovery}},
  booktitle={Proceedings of the AAAI Conference on Artificial Intelligence (AAAI)},
  year={2023}
}

@inproceedings{mcmahan2017communication,
  title={{Communication-Efficient Learning of Deep Networks From Decentralized Data}},
  author={McMahan, Brendan and Moore, Eider and Ramage, Daniel and Hampson, Seth and y Arcas, Blaise Aguera},
  booktitle={{Proceedings of the International Conference on Artificial Intelligence and Statistics (AISTATS)}},
  year={2017}
}

@article{fan2020hybridbfl,
  title={Hybrid blockchain-based resource trading system for federated learning in edge computing},
  author={Fan, Sizheng and Zhang, Hongbo and Zeng, Yuchen and Cai, Wei},
  journal={IEEE Internet of Things Journal},
  volume={8},
  number={4},
  pages={2252--2264},
  year={2020},
  publisher={IEEE}
}

@inproceedings{mugunthan2021blockflow,
  title={BlockFLow: Decentralized, privacy-preserving, and accountable federated machine learning},
  author={Mugunthan, Vaikkunth and Rahman, Ravi and Kagal, Lalana},
  booktitle={International Congress on Blockchain and Applications},
  pages={233--242},
  year={2021},
  organization={Springer}
}

@article{ying2024bitfl,
  title={BIT-FL: Blockchain-enabled incentivized and secure federated learning framework},
  author={Ying, Chenhao and Xia, Fuyuan and Wei, David SL and Yu, Xinchun and Xu, Yibin and Zhang, Weiting and Jiang, Xikun and Jin, Haiming and Luo, Yuan and Zhang, Tao and others},
  journal={IEEE Transactions on Mobile Computing},
  year={2024},
  publisher={IEEE}
}

@article{lin2024refiner,
  title={Refiner: a reliable and efficient incentive-driven federated learning system powered by blockchain},
  author={Lin, Hong and Chen, Ke and Jiang, Dawei and Shou, Lidan and Chen, Gang},
  journal={The VLDB Journal},
  volume={33},
  number={3},
  pages={807--831},
  year={2024},
  publisher={Springer}
}

@inproceedings{he2016deep,
  title={Deep residual learning for image recognition},
  author={He, Kaiming and Zhang, Xiangyu and Ren, Shaoqing and Sun, Jian},
  booktitle={Proceedings of the IEEE conference on computer vision and pattern recognition},
  year={2016}
}

@article{krizhevsky2009learning,
  title={Learning multiple layers of features from tiny images},
  author={Krizhevsky, Alex and Hinton, Geoffrey and others},
  year={2009},
  publisher={Toronto, ON, Canada}
}

@inproceedings{zhang2021incentive,
  title={{Incentive Mechanism for Horizontal Federated Learning Based on Reputation and Reverse Auction}},
  author={Zhang, Jingwen and Wu, Yuezhou and Pan, Rong},
  booktitle={Proceedings of the ACM on Web Conference (WWW)},
  year={2021}
}

@inproceedings{sun2024hifi,
  title={{HiFi-Gas: Hierarchical Federated Learning Incentive Mechanism Enhanced Gas Usage Estimation}},
  author={Sun, Hao and Tang, Xiaoli and Yang, Chengyi and Yu, Zhenpeng and Wang, Xiuli and Ding, Qijie and Li, Zengxiang and Yu, Han},
  booktitle={Proceedings of the AAAI Conference on Artificial Intelligence (AAAI)},
  year={2024}
}

@inproceedings{wang2025dealing,
  title={Dealing with Noisy Data in Federated Learning: An Incentive Mechanism with Flexible Pricing},
  author={Wang, Hengzhi and Chen, Haoran and Ma, Minghe and Cui, Laizhong},
  booktitle={Proceedings of the ACM on Web Conference (WWW)},
  year={2025}
}

@article{wood2014ethereum,
  title={{Ethereum: A Secure Decentralised Generalised Transaction Ledger}},
  author={Wood, Gavin and others},
  journal={Ethereum project yellow paper},
  volume={151},
  number={2014},
  pages={1--32},
  year={2014}
}

@article{shapley1953value,
  title={{A Value for N-Person Games}},
  author={SHAPLEY, LS},
  journal={Contributions to the Theory of Games},
  pages={307--317},
  year={1953},
}

@article{wang2020principled,
  title={{A Principled Approach to Data Valuation for Federated Learning}},
  author={Wang, Tianhao and Rausch, Johannes and Zhang, Ce and Jia, Ruoxi and Song, Dawn},
  journal={Federated Learning: Privacy and Incentive},
  pages={153--167},
  year={2020},
}

@inproceedings{li2020federatedmlsys,
  title={{Federated Optimization in Heterogeneous Networks}},
  author={Li, Tian and Sahu, Anit Kumar and Zaheer, Manzil and Sanjabi, Maziar and Talwalkar, Ameet and Smith, Virginia},
  booktitle    = {Proceedings of the Conference on Machine Learning and Systems (MLSys)},
  year         = {2020},
}

@inproceedings{khanezaei2014framework,
  title={{A Framework Based on RSA and AES Encryption Algorithms for Cloud Computing Services}},
  author={Khanezaei, Nasrin and Hanapi, Zurina Mohd},
  booktitle={Proceedings of the Conference on Systems, Process and Control (ICSPC)},
  year={2014},
}

@article{behera2021federated,
  title={{Federated Learning Using Smart Contracts on Blockchains, Based on Reward Driven Approach}},
  author={Behera, Monik Raj and Upadhyay, Sudhir and Shetty, Suresh},
  journal={arXiv preprint arXiv:2107.10243},
  year={2021}
}

@inproceedings{gupta2002performance,
  title={{Performance Analysis of Elliptic Curve Cryptography for SSL}},
  author={Gupta, Vipul and Gupta, Sumit and Chang, Sheueling and Stebila, Douglas},
  booktitle={Proceedings of the ACM Workshop on Wireless Security (WiSE)},
  year={2002}
}

@article{hsu2019measuring,
      title={{Measuring the Effects of Non-Identical Data Distribution for Federated Visual Classification}}, 
      author={Hsu, Tzu-Ming Harry and Qi, Hang and Brown, Matthew},
      journal={arXiv preprint arXiv:1909.06335},
      year={2019}
}

@inproceedings{chen2024contribution,
  author       = {Yiwei Chen and
                  Kaiyu Li and
                  Guoliang Li and
                  Yong Wang},
  title        = {Contributions Estimation in Federated Learning: {A} Comprehensive
                  Experimental Evaluation},
booktitle = {Proceedings of the VLDB Endowment},
year = {2024},
}

@inproceedings{gao2021fifl,
author = {Gao, Liang and Li, Li and Chen, Yingwen and Zheng, Wenli and Xu, ChengZhong and Xu, Ming},
title = {FIFL: A Fair Incentive Mechanism for Federated Learning},
year = {2021},
booktitle = {Proceedings of the 50th International Conference on Parallel Processing}
}

@article{wang2023incentive,
  author       = {Zhilin Wang and
                  Qin Hu and
                  Ruinian Li and
                  Minghui Xu and
                  Zehui Xiong},
  title        = {Incentive Mechanism Design for Joint Resource Allocation in Blockchain-Based
                  Federated Learning},
  journal      = {{IEEE} Trans. Parallel Distributed Syst.},
  volume       = {34},
  number       = {5},
  pages        = {1536--1547},
  year         = {2023},
}

@inproceedings{wang2024fast,
  title={Fast, robust and interpretable participant contribution estimation for federated learning},
  author={Wang, Yong and Li, Kaiyu and Luo, Yuyu and Li, Guoliang and Guo, Yunyan and Wang, Zhuo},
  booktitle={2024 IEEE 40th International Conference on Data Engineering (ICDE)},
  year={2024}
}

@article{tang2025game,
  title={Game-Theoretic Incentive Mechanism for Blockchain-Based Federated Learning},
  author={Tang, Wenzheng and Liu, Erwu and Ni, Wei and Qu, Xinyu and Huang, Butian and Li, Kezhi and Niyato, Dusit and Jamalipour, Abbas},
  journal={IEEE Transactions on Mobile Computing},
  year={2025},
  publisher={IEEE}
}

@inproceedings{zheng2023secure,
  author       = {Shuyuan Zheng and
                  Yang Cao and
                  Masatoshi Yoshikawa},
  title        = {Secure Shapley Value for Cross-Silo Federated Learning},
  year         = {2023},
booktitle = {Proceedings of the VLDB Endowment},
}

@inproceedings{wei2025efficient,
  author       = {Shuyue Wei and
                  Yongxin Tong and
                  Zimu Zhou and
                  Tianran He and
                  Yi Xu},
  title        = {Efficient Data Valuation Approximation in Federated Learning: {A}
                  Sampling-Based Approach},
  booktitle    = {41st {IEEE} International Conference on Data Engineering, {ICDE}},
  year         = {2025},
}

@misc{netsuite_crossborder_2025,
  title        = {Cross-Border Payments: Types, Advantages, and Challenges},
  author       = {{NetSuite}},
  year         = {2025},
  howpublished = {\url{https://www.netsuite.com/portal/resource/articles/accounting/cross-border-payments.shtml}},
  note         = {Accessed: 2026-02-09}
}

@misc{worldbank_remittance_2024,
  title        = {Remittance Prices Worldwide},
  author       = {{World Bank}},
  year         = {2024},
  howpublished = {\url{https://remittanceprices.worldbank.org}},
  note         = {Accessed: 2026-02-09}
}

@news{reutersG202025,
  title   = {G20's Cross-Border Payments Push Set to Miss 2027 Target},
  author  = {Staff, Reuters},
  journal = {Reuters},
  year    = {2025},
  url     = {https://www.reuters.com/business/retail-consumer/g20s-cross-border-payments-push-set-miss-2027-target-2025-10-09/},
  note    = {Accessed: 2026-02-09}
}

@misc{ethereum_stats_2025,
  title        = {Ethereum Network Statistics},
  author       = {{PatentPC}},
  year         = {2025},
  howpublished = {\url{https://patentpc.com/blog/ethereum-network-growth-gas-fees-staking-usage-stats}},
  note         = {Accessed: 2026-02-09}
}

@misc{ethereum_gas_fees_2025,
  title        = {Ethereum Gas Fees Statistics},
  author       = {{SQ Magazine}},
  year         = {2025},
  howpublished = {\url{https://sqmagazine.co.uk/ethereum-gas-fees-statistics/}},
  note         = {Accessed: 2026-02-09}
}

@article{he2020fedml,
  title={Fedml: A research library and benchmark for federated machine learning},
  author={He, Chaoyang and Li, Songze and So, Jinhyun and Zeng, Xiao and Zhang, Mi and Wang, Hongyi and Wang, Xiaoyang and Vepakomma, Praneeth and Singh, Abhishek and Qiu, Hang and others},
  journal={arXiv preprint arXiv:2007.13518},
  year={2020}
}

@article{beutel2020flower,
  title={Flower: A friendly federated learning research framework},
  author={Beutel, Daniel J and Topal, Taner and Mathur, Akhil and Qiu, Xinchi and Fernandez-Marques, Javier and Gao, Yan and Sani, Lorenzo and Li, Kwing Hei and Parcollet, Titouan and de Gusm{\~A}{\c{G}}o, Pedro Porto Buarque and others},
  journal={arXiv preprint arXiv:2007.14390},
  year={2020}
}

@article{reina2021openfl,
  title={OpenFL: An open-source framework for Federated Learning},
  author={Reina, G Anthony and Gruzdev, Alexey and Foley, Patrick and Perepelkina, Olga and Sharma, Mansi and Davidyuk, Igor and Trushkin, Ilya and Radionov, Maksim and Mokrov, Aleksandr and Agapov, Dmitry and others},
  journal={arXiv preprint arXiv:2105.06413},
  year={2021}
}

@book{reuter2025payment,
  title={Payment Frictions, Capital Flows, and Exchange Rates},
  author={Reuter, Marco and Agur, Mr Itai and Copestake, Alexander and Peria, Maria Soledad Martinez and Teoh, Mr Ken},
  year={2025},
  publisher={International Monetary Fund}
}

@book{bindseil2022towards,
  title={Towards the holy grail of cross-border payments},
  author={Bindseil, Ulrich and Pantelopoulos, George},
  number={2693},
  year={2022},
  publisher={ECB Working Paper}
}

@article{alhaidari2025chain,
  title={On-Chain Decentralized Learning and Cost-Effective Inference for DeFi Attack Mitigation},
  author={Alhaidari, Abdulrahman and Palanisamy, Balaji and Krishnamurthy, Prashant},
  journal={arXiv preprint arXiv:2510.16024},
  year={2025}
}

@article{kang2019incentive,
  title={Incentive Mechanism for Reliable Federated Learning: A joint optimization approach to combining reputation and contract theory},
  author={Kang, Jiawen and Xiong, Zehui and Niyato, Dusit and Xie, Shengli and Zhang, Junshan},
  journal={IEEE Internet of Things Journal},
  volume={6},
  number={6},
  pages={10700--10714},
  year={2019},
  publisher={IEEE}
}

@inproceedings{bonawitz2017practical,
  title={Practical secure aggregation for privacy-preserving machine learning},
  author={Bonawitz, Keith and Ivanov, Vladimir and Kreuter, Ben and Marcedone, Antonio and McMahan, H Brendan and Patel, Sarvar and Ramage, Daniel and Segal, Aaron and Seth, Karn},
  booktitle={proceedings of the 2017 ACM SIGSAC Conference on Computer and Communications Security},
  year={2017}
}

@article{yu2023ironforge,
  title={IronForge: An open, secure, fair, decentralized federated learning},
  author={Yu, Guangsheng and Wang, Xu and Sun, Caijun and Wang, Qin and Yu, Ping and Ni, Wei and Liu, Ren Ping},
  journal={IEEE Transactions on Neural Networks and Learning Systems},
  year={2023},
  publisher={IEEE}
}

@article{roth2022nvidia,
  title={Nvidia flare: Federated learning from simulation to real-world},
  author={Roth, Holger R and Cheng, Yan and Wen, Yuhong and Yang, Isaac and Xu, Ziyue and Hsieh, Yuan-Ting and Kersten, Kristopher and Harouni, Ahmed and Zhao, Can and Lu, Kevin and others},
  journal={arXiv preprint arXiv:2210.13291},
  year={2022}
}

@article{galtier2019substra,
  title={Substra: a framework for privacy-preserving, traceable and collaborative machine learning},
  author={Galtier, Mathieu N and Marini, Camille},
  journal={arXiv preprint arXiv:1910.11567},
  year={2019}
}

@inproceedings{li2025veryfl,
  title={{VeryFL: A Verify Federated Learning Framework Embedded With Blockchain}},
  author={Li, Yihao and Lai, Yanyi and Li, Xiaoli and Chen, Chuan},
  booktitle={International Conference on Blockchain and Trustworthy Systems},
  year={2025},
}

@article{liu2022gtg,
  title={GTG-Shapley: Efficient and Accurate Participant Contribution Evaluation in Federated Learning},
  author={Liu, Zelei and Chen, Yuanyuan and Yu, Han and Liu, Yang and Cui, Lizhen},
  journal={ACM Transactions on intelligent Systems and Technology (TIST)},
  volume={13},
  number={4},
  pages={1--21},
  year={2022},
  publisher={ACM New York, NY}
}

@inproceedings{reddi2021adaptive,
  title={Adaptive Federated Optimization},
  author={Reddi, Sashank J and Charles, Zachary and Zaheer, Manzil and Garrett, Zachary and Rush, Keith and Kone{\v{c}}n{\`y}, Jakub and Kumar, Sanjiv and McMahan, Hugh Brendan},
  booktitle={International Conference on Learning Representations},
  year={2021}
}

@inproceedings{xu2021gradient,
  title={Gradient driven rewards to guarantee fairness in collaborative machine learning},
  author={Xu, Xinyi and Lyu, Lingjuan and Ma, Xingjun and Miao, Chenglin and Foo, Chuan Sheng and Low, Bryan Kian Hsiang},
  booktitle={Advances in Neural Information Processing Systems},
  year={2021}
}

@article{zhan2021survey,
  title={A survey of incentive mechanism design for federated learning},
  author={Zhan, Yufeng and Zhang, Jie and Hong, Zicong and Wu, Leijie and Li, Peng and Guo, Song},
  journal={IEEE Transactions on Emerging Topics in Computing},
  volume={10},
  number={2},
  pages={1035--1044},
  year={2021},
  publisher={IEEE}
}

@article{kairouz2021advances,
  title={Advances and open problems in federated learning},
  author={Kairouz, Peter and McMahan, H Brendan},
  journal={Foundations and trends in machine learning},
  volume={14},
  number={1-2},
  pages={1--210},
  year={2021},
  publisher={Emerald Publishing Limited}
}

@inproceedings{yu2020fairness,
  title={A fairness-aware incentive scheme for federated learning},
  author={Yu, Han and Liu, Zelei and Liu, Yang and Chen, Tianjian and Cong, Mingshu and Weng, Xi and Niyato, Dusit and Yang, Qiang},
  booktitle={Proceedings of the AAAI/ACM Conference on AI, Ethics, and Society},
  year={2020}
}

@article{weng2019deepchain,
  title={Deepchain: Auditable and privacy-preserving deep learning with blockchain-based incentive},
  author={Weng, Jiasi and Weng, Jian and Zhang, Jilian and Li, Ming and Zhang, Yue and Luo, Weiqi},
  journal={IEEE Transactions on Dependable and Secure Computing},
  volume={18},
  number={5},
  pages={2438--2455},
  year={2019},
  publisher={IEEE}
}

@inproceedings{zhu2019deep,
  title={Deep leakage from gradients},
  author={Zhu, Ligeng and Liu, Zhijian and Han, Song},
  booktitle={Advances in Neural Information Processing Systems},
  year={2019}
}

@article{nguyen2021federated,
  title={Federated learning meets blockchain in edge computing: Opportunities and challenges},
  author={Nguyen, Dinh C and Ding, Ming and Pham, Quoc-Viet and Pathirana, Pubudu N and Le, Long Bao and Seneviratne, Aruna and Li, Jun and Niyato, Dusit and Poor, H Vincent},
  journal={IEEE Internet of Things Journal},
  volume={8},
  number={16},
  pages={12806--12825},
  year={2021},
  publisher={IEEE}
}

@inproceedings{melis2019exploiting,
  title={Exploiting unintended feature leakage in collaborative learning},
  author={Melis, Luca and Song, Congzheng and De Cristofaro, Emiliano and Shmatikov, Vitaly},
  booktitle={2019 IEEE symposium on security and privacy (SP)},
  year={2019}
}

@inproceedings{blanchard2017machine,
  title={Machine learning with adversaries: Byzantine tolerant gradient descent},
  author={Blanchard, Peva and El Mhamdi, El Mahdi and Guerraoui, Rachid and Stainer, Julien},
  booktitle={Advances in Neural Information Processing Systems},
  year={2017}
}

@inproceedings{bagdasaryan2020backdoor,
  title={How to backdoor federated learning},
  author={Bagdasaryan, Eugene and Veit, Andreas and Hua, Yiqing and Estrin, Deborah and Shmatikov, Vitaly},
  booktitle={International conference on artificial intelligence and statistics},
  year={2020}
}

@inproceedings{fraboni2021free,
  title={Free-rider attacks on model aggregation in federated learning},
  author={Fraboni, Yann and Vidal, Richard and Lorenzi, Marco},
  booktitle={International conference on artificial intelligence and statistics},
  year={2021}
}

@article{xu2019verifynet,
  title={VerifyNet: Secure and verifiable federated learning},
  author={Xu, Guowen and Li, Hongwei and Liu, Sen and Yang, Kan and Lin, Xiaodong},
  journal={IEEE Transactions on Information Forensics and Security},
  volume={15},
  pages={911--926},
  year={2019},
  publisher={IEEE}
}

@inproceedings{bell2020secure,
  title={Secure single-server aggregation with (poly) logarithmic overhead},
  author={Bell, James Henry and Bonawitz, Kallista A and Gasc{\'o}n, Adri{\`a} and Lepoint, Tancr{\`e}de and Raykova, Mariana},
  booktitle={Proceedings of the 2020 ACM SIGSAC conference on computer and communications security},
  year={2020}
}

@article{wahrstatter2024openfl,
  title={OpenFL: A scalable and secure decentralized federated learning system on the Ethereum blockchain},
  author={Wahrst{\"a}tter, Anton and Khan, Sajjad and Svetinovic, Davor},
  journal={Internet of Things},
  volume={26},
  pages={101174},
  year={2024},
  publisher={Elsevier}
}
